\documentclass[PAPER, atlasdraft=false, texlive=2023, UKenglish, cernpreprint=true, coverpage=false, texmf, orcidlogo]{atlasdoc} %
\usepackage[subfigure=true]{atlaspackage}
\usepackage{atlasbiblatex}

\usepackage{atlasphysics}
\graphicspath{{logos/}{figures/}}

\usepackage{ANA-HDBS-2019-27-PAPER-defs}

\usepackage{array}
\usepackage[super]{nth} %
\usepackage{pifont} %
\newcolumntype{M}[1]{>{\centering\arraybackslash}m{#1}}
\usepackage{arydshln}
\usepackage{multirow}
\usepackage{caption}

\AtlasTitle{Search for the non-resonant production of Higgs boson pairs via gluon fusion and vector-boson fusion in the $b\bar{b}\tau^+\tau^-$ final state in proton--proton collisions at $\sqrt{s} = 13$~TeV with the ATLAS detector}

\AtlasAbstract{%
A search for the non-resonant production of Higgs boson pairs in the $HH\rightarrow b\bar{b}\tau^+\tau^-$ channel is performed using 140~fb$^{-1}$ of proton--proton collisions at a centre-of-mass energy of $13$~TeV recorded by the ATLAS detector at the CERN Large Hadron Collider.
The analysis strategy is optimised to probe anomalous values of the Higgs boson self-coupling modifier $\kappa_\lambda$ and of the quartic $HHVV$ ($V = W,Z$) coupling modifier $\kappa_{2V}$.
No significant excess above the expected background from Standard Model processes is observed.
An observed (expected) upper limit $\mu_{HH}<5.9~(3.3)$ is set at 95\% confidence-level on the Higgs boson pair production cross-section normalised to its Standard Model prediction.
The coupling modifiers are constrained to an observed (expected) 95\% confidence interval of $-3.1 < \kappa_\lambda < 9.0$ ($-2.5 < \kappa_\lambda < 9.3$) and $-0.5 < \kappa_{2V} < 2.7$ ($-0.2 < \kappa_{2V} < 2.4$), assuming all other Higgs boson couplings are fixed to the Standard Model prediction.
The results are also interpreted in the context of effective field theories via constraints on anomalous Higgs boson couplings and Higgs boson pair production cross-sections assuming different kinematic benchmark scenarios. %
}
\AtlasRefCode{HDBS-2019-27}

\PreprintIdNumber{CERN-EP-2024-091}

\AtlasJournalRef{\PRD 110 (2024) 032012}
\AtlasDOI{10.1103/PhysRevD.110.032012}

\hypersetup{pdftitle={ATLAS document},pdfauthor={The ATLAS Collaboration}}

\mathchardef\mhyphen="2D
\begin{document}

\maketitle

\tableofcontents

\section{Introduction}

%
%
%
%

%
After the discovery of the Higgs boson ($H$) in 2012~\cite{HIGG-2012-27,CMS-HIG-12-028}, the ATLAS~\cite{ATLAS:2008xda} and CMS~\cite{CMS:2008xjf} Collaborations at the Large Hadron Collider~\cite{Evans:2008zzb} (LHC) have been focusing on the measurement of its properties~\cite{HIGG-2013-17-witherratum,HIGG-2019-01,CMS-HIG-19-009,HIGG-2018-32,CMS-HIG-21-013,ATLAS:2022vkf,CMS:2022dwd}.
All features probed so far for this new particle are consistent with the Standard Model (SM) predictions~\cite{PhysRevLett.13.321,HIGGS1964132,PhysRevLett.13.508,PhysRevLett.13.585,PhysRev.145.1156,PhysRev.155.1554} for a Higgs boson with an observed mass $m_{H}$ near 125~GeV~\cite{PhysRevLett.114.191803,HIGG-2022-20-journal}.

In this extensive measurement programme, the properties of interactions involving multiple Higgs bosons remain to be verified.
In the SM, these can be characterised by the trilinear and quartic self-couplings $\lhhh$ and $\lhhhh$, which are both equal to the coefficient $\lambda$ of the quartic term of the Higgs field potential $V(\phi) = \mu^2|\phi|^2 + \lambda|\phi|^4$.
The quartic couplings $g_{HHVV}$ ($V = W, Z$) characterise the interactions between two Higgs bosons and two $W$ or $Z$ bosons, and are related in the SM to the $HWW$ and $HZZ$ couplings $g_{HVV}$ through the relation $g_{HHVV} = g_{HVV}/2\nu$, where $\nu$ is the vacuum expectation value of the Higgs field.
Significant deviations from the SM predictions for these couplings would provide a strong sign of beyond the Standard Model (BSM) physics~\cite{PhysRevD.101.075023}.

The most sensitive test of Higgs boson self-interactions comes from processes of Higgs boson pair production (\HH) such as gluon fusion (ggF) and vector-boson fusion (VBF) \HH production. %
Measuring the cross-section of these processes offers a direct probe of the values of these couplings, through their scale factors with respect to the SM predictions: $\kl = \lhhh/\lhhh^{SM}$, affecting both ggF and VBF production, and $\kv = g_{HVV}/g^{SM}_{HVV}$ and $\kvv = g_{HHVV}/g_{HHVV}^{SM}$, which only impact VBF production.

\begin{figure}[!tbh]
\centering
\subfigure[]{\includegraphics[width=0.45\textwidth]{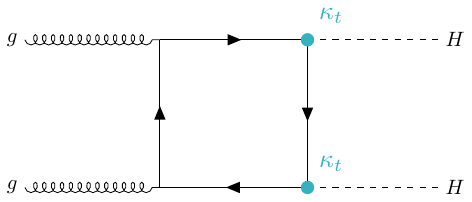}} %
~
\subfigure[]{\includegraphics[width=0.45\textwidth]{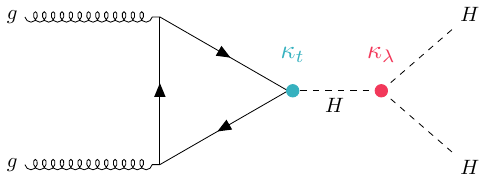}} %
\caption{Leading-order Feynman diagrams of the ggF \HH production process: (a) box and (b) triangle diagrams.
The Higgs self-coupling modifier is shown as \kl, while the modifier for the coupling of the Higgs boson to the top quark is shown as \kappat.}
\label{fig:ggFdiagrams}
\end{figure}

\begin{figure}[!tbh]
\centering
\subfigure[]{\includegraphics[width=0.30\textwidth]{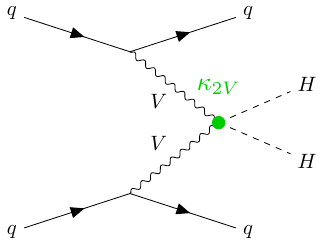} \label{sfg:vvhh}}
~
\subfigure[]{\includegraphics[width=0.30\textwidth]{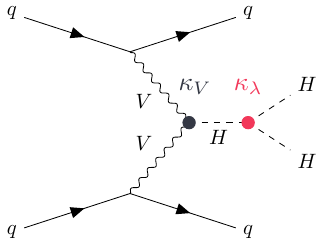}}
~
\subfigure[]{\includegraphics[width=0.30\textwidth]{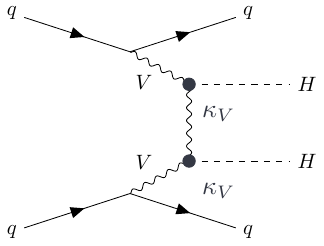}\label{sfg:vvh}}
\caption{Leading-order Feynman diagrams of the VBF \HH production process.
The $g_{HHVV}$ coupling modifier is shown as \kvv, the Higgs self-coupling modifier as \kl, and the modifier for the coupling of the Higgs boson to the SM vector bosons $V$ as \kv.}
\label{fig:VBFdiagrams}
\end{figure}

The dominant mode for \HH production is ggF, with a cross-section of $\sigma_{\textup{ggF}}^{\textup{SM}} = 31.1^{+2.1}_{-7.2}$ fb~\cite{Dawson:1998py,Borowka:2016ehy,Baglio:2018lrj,deFlorian:2013jea,Shao:2013bz,deFlorian:2015moa,Grazzini:2018bsd,Baglio:2020wgt}, calculated at next-to-next-to-leading order (NNLO) including finite top-quark-mass effects for $m_{H} = 125$~GeV at $\sqrt{s} = 13$~TeV, resulting from the destructive interference between the leading-order (LO) Feynman diagrams shown in Figure~\ref{fig:ggFdiagrams}.
The second most common \HH production mechanism at the LHC is VBF, with a total cross-section of $\sigma_{\textup{VBF}}^{\textup{SM}} = 1.73 \pm 0.04$ fb~\cite{Baglio:2012np,Frederix:2014hta,Ling:2014sne,Dreyer:2018rfu,Dreyer:2018qbw} calculated at next-to-next-to-next-to-leading order (N$^3$LO) for $m_{H} = 125$~GeV at $\sqrt{s} = 13$~TeV.
The LO diagrams for this process are shown in Figure~\ref{fig:VBFdiagrams}.

The most stringent constraints on SM \HH production by the ATLAS Collaboration were obtained through a statistical combination~\cite{HDBS-2022-03} of the results in the \bbyy~\cite{HDBS-2018-34}, \bbtt~\cite{HDBS-2018-40} and \bbbb~\cite{HDBS-2019-29} final states, exploiting the entire sample of proton--proton ($pp$) collisions provided by the LHC during its second phase (Run~2, 2015--2018).
The 95\% confidence level (CL) observed (expected) upper limit on the \HH production signal strength $\mu_{HH} = (\sigma_{\textup{ggF}} + \sigma_{\textup{VBF}})/(\sigma_{\textup{ggF}}^{\textup{SM}} + \sigma_{\textup{VBF}}^{\textup{SM}})$ from the combination is 2.4~(2.9) with respect to the case $\mu_{HH} = 0$, which is later referred to as the \emph{background-only hypothesis}.
Using the values of the combination test statistic as a function of either \kl or \kvv, when all other coupling modifiers are set to unity, the observed (expected with respect to the SM prediction) 95\% confidence intervals (CIs) are $\kl \in [-0.6, 6.6]$~($[-2.1, 7.8]$) and $\kvv\in [0.1, 2.0]$~($[0.0, 2.1]$).
The CMS Collaboration also performed a combination of \HH analyses in multiple final states, based on their full Run~2 dataset~\cite{CMS:2022dwd}.
The observed (expected) 95\% CL upper limit on $\mu_{HH}$ is 3.4~(2.5).
The observed allowed ranges of \kl and \kvv are restricted to $\kl\in [-1.24, 6.49]$ and $\kvv\in [0.67, 1.38]$.
No significant excess over the background-only hypothesis is observed by any of these analyses.

The \bbtt final state captures 7.3\% of all \HH final states and provides a compromise between expected signal event yield and background contamination.
This leads to a sensitivity similar to that of the \bbbb and \bbyy decay modes.
The latest ATLAS results for $HH\rightarrow\bbtt$ with the full LHC Run~2 dataset are documented in Ref.~\cite{HDBS-2018-40}. %
They result in an observed (expected) 95\% CL upper limit on the total \HH production cross-section of 4.7~(3.9) times the SM prediction with respect to the background-only hypothesis.
Moreover, observed~(expected) 95\% CIs on \kl and \kvv are $\kl \in [-2.7, 9.5]$~($[-3.1, 10.2]$) and $\kvv\in [-0.6, 2.7]$~($[-0.5, 2.7]$), respectively~\cite{HDBS-2022-03}.
Recent results from the CMS Collaboration~\cite{CMS-HIG-20-010} set an observed (expected) upper limit of 3.3~(5.2) times the SM production cross-section at 95\% CL over the background-only hypothesis.
Additionally, constraints are set on \kl and \kvv with respect to the background-only hypothesis, leading to the observed (expected) 95\% CL constraints $\kl\in [-1.7, 8.7]$~($[-2.9, 9.8]$) and $\kvv\in[-0.4, 2.6]$~($[-0.6, 2.8]$).

In this paper, an updated search for non-resonant Higgs boson pair production in the \bbtt final state with the full Run~2 ATLAS dataset is presented, superseding and expanding upon the
non-resonant results of Ref.~\cite{HDBS-2018-40}.
Compared with the previous publication, the overall object identification, trigger strategy and event selection in the signal-enriched regions is unchanged, but an optimised classification of the selected events is implemented to enhance the sensitivity to \kl and to the VBF production mode.
Improved multivariate classifiers are used to exploit the kinematic features of SM VBF \HH production to define a dedicated VBF category, which improves the sensitivity to anomalous values of the coupling modifiers \kl and \kvv.
Updated Monte Carlo (MC) predictions are used for describing the main backgrounds of top-quark pair production ($t\bar{t}$) and $Z$ boson production in association with heavy-flavour quarks, leading to a more accurate modelling of these processes and enhancing the statistical power of the simulation.
The event selection for the auxiliary measurement of the background from $Z$ boson production in association with heavy-flavour quarks is adapted to improve the consistency of the kinematic properties of this process with the signal-enriched regions.
Finally, the obtained results are reinterpreted in the context of effective field theories (EFTs).
These changes in analysis set-up and strategy lead to an improvement of the expected 95\% CL limit on $\mu_{HH}$ of 15\%.
The widths of the 95\% CIs on \kl and \kvv with respect to the SM expectation are reduced by 11\% and 19\%, respectively, compared with the analysis described in Ref.~\cite{HDBS-2018-40}.

This paper is organised as follows. %
Section~\ref{sec:detector} describes the ATLAS detector.
The collected dataset and the simulated event samples are covered in Section~\ref{sec:mc}.
The event selection and categorisation, along with the description of the multivariate classifiers, are described in Section~\ref{sec:esel}.
Section~\ref{sec:systs} lists the systematic uncertainties. %
Finally, Section~\ref{sec:result} summarises the results, Section~\ref{sec:eft} covers the conducted EFT interpretations, and the study is concluded in Section~\ref{sec:concl}.


\section{ATLAS detector} %

\label{sec:detector}
\newcommand{\AtlasCoordFootnote}{%
ATLAS uses a right-handed coordinate system with its origin at the nominal interaction point (IP)
in the centre of the detector and the \(z\)-axis along the beam pipe.
The \(x\)-axis points from the IP to the centre of the LHC ring,
and the \(y\)-axis points upwards.
Polar coordinates \((r,\phi)\) are used in the transverse plane,
\(\phi\) being the azimuthal angle around the \(z\)-axis.
The pseudorapidity is defined in terms of the polar angle \(\theta\) as \(\eta = -\ln \tan(\theta/2)\).
Angular distance is measured in units of \(\Delta R \equiv \sqrt{(\Delta\eta)^{2} + (\Delta\phi)^{2}}\).}
The ATLAS experiment at the LHC is a multipurpose particle detector
with a forward--backward symmetric cylindrical geometry and a near \(4\pi\) coverage in
solid angle.\footnote{\AtlasCoordFootnote}
It consists of an inner tracking detector surrounded by a thin superconducting solenoid
providing a \qty{2}{\tesla} axial magnetic field, electromagnetic and hadron calorimeters, and a muon spectrometer.
The inner tracking detector covers the pseudorapidity range \(|\eta| < 2.5\).
It consists of silicon pixel, silicon microstrip, and transition radiation tracking detectors.
Lead/liquid-argon (LAr) sampling calorimeters provide electromagnetic (EM) energy measurements
with high granularity.
A steel/scintillator-tile hadron calorimeter covers the central pseudorapidity range (\(|\eta| < 1.7\)).
The endcap and forward regions are instrumented with LAr calorimeters
for both the EM and hadronic energy measurements up to \(|\eta| = 4.9\).
The muon spectrometer surrounds the calorimeters and is based on
three large superconducting air-core toroidal magnets with eight coils each.
The field integral of the toroids ranges between \num{2.0} and \qty{6.0}{\tesla\metre}
across most of the detector.
The muon spectrometer includes a system of precision tracking chambers and fast detectors for triggering.
A two-level trigger system is used to select events.
The first-level trigger is implemented in hardware and uses a subset of the detector information
to accept events at a rate below \qty{100}{\kHz}.
This is followed by a software-based trigger that
reduces the accepted event rate to \qty{1}{\kHz} on average
depending on the data-taking conditions.
An extensive software suite~\cite{ATL-SOFT-PUB-2021-001} is used in simulation, in the reconstruction
and analysis of real and simulated events, in detector operations, and in the trigger and data acquisition
systems of the experiment.


\section{Data and simulated event samples}

\label{sec:mc}
The search presented in this paper uses the $pp$ collision dataset collected by the ATLAS experiment during the LHC Run 2 at a centre-of-mass energy of $\sqrt{s} = 13$~TeV.
After data quality requirements~\cite{DAPR-2018-01}, the integrated luminosity of the dataset is $140.1 \pm 1.2$ fb$^{-1}$~\cite{DAPR-2021-01, LUCID2}.

The samples of simulated events used for this study are summarised in Table~\ref{tab:samples}, and they correspond to those already employed in the previous publication~\cite{HDBS-2018-40}, with a few important updates aimed at improving the statistical power of the simulated samples for the main background processes providing a more accurate simulation of the \HH signal.
All generated samples are passed through a detailed simulation of the ATLAS detector response~\cite{SOFT-2010-01} based on \GEANT~\cite{Agostinelli:2002hh}.

The production of $t\bar{t}$ events, and of single-top-quark events in the $Wt-$, $s-$ and $t-$channels is simulated using the \POWHEGBOX[v2]~\cite{Alioli:2010xd} generator together with the \NNPDF[3.0nlo] parton distribution functions (PDF) set~\cite{Ball:2014uwa}.
The simulated events are interfaced to \Pythia[8.230]~\cite{Sjostrand:2014zea} for parton showering and hadronisation using the A14 tune~\cite{ATL-PHYS-PUB-2014-021} together with the \NNPDF[2.3lo]~\cite{Ball:2012cx} PDF set.
The setup of the MC event generator for these production processes is unchanged with respect to Ref.~\cite{HDBS-2018-40}.
For $t\bar{t}$ processes the size of the sample is increased by including two statistically independent sets of events.
The first sample is simulated with the requirement of at least one top quark decaying into a leptonic final state, while in the second sample both top quarks are forced to decay leptonically.
The combination of these samples results in a decrease in the statistical uncertainty related to the $t\bar{t}$ simulation by a factor of approximately two in the selected phase space compared to using only the first sample, as was done in Ref.~\cite{HDBS-2018-40}.

For events containing a $W$ or a $Z$ boson produced in association with jets, new samples are produced with the \Sherpa[2.2.11] generator~\cite{Bothmann:2019yzt}.
Events are simulated using NLO matrix elements for up to two partons, and LO matrix elements for up to five partons calculated using the \OPENLOOPS~\cite{Buccioni:2019sur,Cascioli:2011va,Buccioni:2017yxi,Denner:2016kdg} and Comix~\cite{Gleisberg:2008fv} matrix-element generators.
For these samples, the \NNPDF[3.0nnlo] PDF set is used with dedicated parton shower, matched to the matrix element via the MEPS@NLO prescription~\cite{Hoeche:2012yf}.
The electroweak input scheme is updated compared to the previous version, along with corrected heavy-flavour hadron production fractions.
The \Sherpa[2.2.11] generator allows faster per-event generation time, with significant reduction of the negative weight fraction.
Using a sample generated with \Sherpa[2.2.11] instead of the sample generated with \Sherpa[2.2.1] that was used in Ref.~\cite{HDBS-2018-40} results in a reduction of the MC statistical uncertainty ranging from 30\% to 60\%, depending on the flavour composition of the events and the analysis region.

Diboson ($WW$, $WZ$ and $ZZ$) and $t\bar{t}Z$ production processes are simulated with the \Sherpa[2.2.1] generator, whereas the $t\bar{t}W$ production process is simulated with the \Sherpa[2.2.8] generator.
These samples use the \NNPDF[3.0nnlo] PDF set with dedicated parton shower tuning developed by the \Sherpa authors.
Single Higgs boson production is considered as part of the background in this search, and its production modes are simulated using the \POWHEGBOX[v2] generator and either the \PDFforLHC[15nlo]~\cite{Butterworth:2015oua} with the AZNLO~\cite{STDM-2012-23} tune or the \NNPDF[3.0nlo] PDF sets with the A14 tune.
The setup of the MC event generator for these production processes is unchanged with respect to Ref.~\cite{HDBS-2018-40}.

Signal samples consist of simulated events from non-resonant ggF and VBF production of Higgs boson pairs, with one Higgs boson decaying into \bb and the other one to $\tau^+\tau^-$.
The simulated ggF events are generated with the \POWHEGBOX[v2] generator at NLO with finite top-quark mass, and using the \PDFforLHC[15nlo] PDF set.
The VBF events are generated at LO using the \MGNLO[2.7.3]~\cite{Alwall:2014hca} generator with the \NNPDF[3.0nlo] PDF set.
Parton shower and hadronisation are simulated using \Pythia[8.244] with the A14 tune and the \NNPDF[2.3lo] PDF set.
A reweighting technique based on the particle-level invariant mass \mhh of the Higgs boson pair is applied to the $\kl = 1$ sample to determine the ggF \HH signal yield and kinematic distributions for any value of \kl~\cite{ATL-PHYS-PUB-2019-007}.
The particle-level \mhh spectrum for any generic value of \kl is calculated from the \mhh distributions of three ggF \HH samples generated for $\kl = 0,1$ or $20$.
To determine the potential non-closure in the reweighting process from residual kinematic effects, an additional ggF \HH sample is generated with the same settings as the nominal sample but with the non-SM value of the self-coupling modifier $\kl = 10$, and then passed through detector simulation and reconstruction algorithms.
The reweighting procedure is validated by comparing the predicted event yields and kinematic distributions of the simulated sample generated with $\kl = 1$ and reweighted to $\kl = 10$ with those of the simulated sample generated under the hypothesis $\kl = 10$.
Furthermore, 12 additional VBF \HH samples are generated and simulated with the same setup as the nominal VBF sample, but using non-SM combinations of the coupling modifiers \kl, \kvv and \kv.
A basis for a linear combination is formed by the SM sample and five of the other 12 samples, corresponding to the combinations of the \kl, \kvv and \kv couplings (1, 1.5, 1), (0, 1, 1), (10, 1, 1), (1, 3, 1), ($-5$, 1, 0.5).
This approach is used to determine the expected VBF \HH yields and distributions for any value of \kl, \kvv and \kv.
The remaining seven samples are compared to the corresponding predictions from the interpolation procedure for validation purposes.
The same procedure was used in the measurements presented in Ref.~\cite{HDBS-2022-03}. %
For both the ggF and the VBF production mode, good closure between the simulated and reweighted samples for alternative \kl, \kv and \kvv values is observed within statistical uncertainties.

The Higgs boson mass is assumed to be 125~GeV in both the simulation and the analysis of the data.
All samples are normalised to the same cross-section calculations detailed in Ref~\cite{HDBS-2018-40}.
The impacts of the differences with respect to the best-fit values of the $m_H$ measurements reported in Refs.~\cite{PhysRevLett.114.191803,HIGG-2022-20-journal} and the effects of the corresponding experimental uncertainties on $m_H$, are negligible.

\begin{table}[tbhp!]
\captionsetup{skip=2pt}
\caption{Generators used to simulate the signal and background processes.
If not otherwise specified, the order of the cross-section calculation refers to the expansion in the strong coupling constant ($\alpha_S$).
The acronyms ME, PS and UE are used for matrix element, parton shower and underlying event, respectively.
More details on the simulation of the signal and background samples are described in the text and in Ref.~\cite{HDBS-2018-40}.
($\dagger$) The NNLO(QCD)+NLO(EW) cross-section calculation for the $pp\rightarrow ZH$ process already includes the $gg\rightarrow ZH$ contribution.
The $qq\rightarrow ZH$ process is normalised to the NNLO(QCD)+NLO(EW) cross-section for the $pp\rightarrow ZH$ process, after subtracting the $gg\rightarrow ZH$ contribution.
\protect\label{tab:samples}}
\begin{center}{\small
\resizebox{\textwidth}{!}{\begin{tabular}{lllllll}
\toprule
\hspace*{0.3cm} Process & ME generator & \shortstack{ME QCD\\order} & PDF set & \shortstack{PS and\\hadronisation} & \shortstack{UE model\\tune} & \shortstack{Cross-section\\order} \\
\midrule
\multicolumn{5}{l}{\BF{Signal}} \\
\hspace*{0.3cm} $gg\rightarrow HH$ (ggF)   &\POWHEGBOX[v2]~\cite{Alioli:2010xd} & NLO & \PDFforLHC[15nlo]~\cite{Butterworth:2015oua} & \Pythia[8.244]~\cite{Sjostrand:2014zea} & A14~\cite{ATL-PHYS-PUB-2014-021} & NNLO FTApprox \\
\hspace*{0.3cm} $qq\rightarrow qqHH$ (VBF)   & \MGNLO[2.7.3]~\cite{Alwall:2014hca} & LO & \NNPDF[3.0nlo]~\cite{Ball:2014uwa} & \Pythia[8.244] & A14 & N$^3$LO(QCD) \\
\midrule
\multicolumn{5}{l}{\BF{Top-quark}} \\
\hspace*{0.3cm} $t\bar{t}$  &\POWHEGBOX[v2] & NLO & \NNPDF[3.0nlo] & \Pythia[8.230] & A14 & NNLO+NNLL\\
\hspace*{0.3cm} $t$-channel &\POWHEGBOX[v2] & NLO & \NNPDF[3.0nlo] & \Pythia[8.230] & A14 & NLO \\
\hspace*{0.3cm} $s$-channel &\POWHEGBOX[v2] & NLO & \NNPDF[3.0nlo] & \Pythia[8.230] & A14 & NLO \\
\hspace*{0.3cm} $Wt$        &\POWHEGBOX[v2] & NLO & \NNPDF[3.0nlo] & \Pythia[8.230] & A14 & NLO \\
\hspace*{0.3cm} $t\bar{t}Z$ & \Sherpa[2.2.1]~\cite{Bothmann:2019yzt} & NLO & \NNPDF[3.0nnlo] &  \Sherpa[2.2.1] & - & NLO \\
\hspace*{0.3cm} $t\bar{t}W$ & \Sherpa[2.2.8] & NLO & \NNPDF[3.0nnlo] &  \Sherpa[2.2.8] & - & NLO \\
\midrule
\multicolumn{5}{l}{\BF{Vector boson + jets}} \\
\hspace*{0.3cm} $W/Z+$jets & \Sherpa[2.2.11] & \shortstack{NLO($\leq 2$ jets)\\ LO($3,4$ jets)} & \NNPDF[3.0nnlo] &  \Sherpa[2.2.11] & - & NNLO \\
\midrule
\multicolumn{5}{l}{\BF{Diboson}} \\
\hspace*{0.3cm}$WW,WZ,ZZ$ & \Sherpa[2.2.1] & \shortstack{NLO($\leq 1$ jets) \\ LO($2,3$ jets)} & \NNPDF[3.0nnlo] &  \Sherpa[2.2.1] & - & NLO \\
\midrule
\multicolumn{5}{l}{\BF{Single Higgs boson}} \\
\hspace*{0.3cm} ggF         &\POWHEGBOX[v2] & NNLO & \PDFforLHC[15nnlo] &  \Pythia[8.212] & AZNLO~\cite{STDM-2012-23} & N$^3$LO(QCD)+NLO(EW) \\
\hspace*{0.3cm} VBF         &\POWHEGBOX[v2] & NLO & \PDFforLHC[15nlo] &  \Pythia[8.212] & AZNLO & NNLO(QCD)+NLO(EW) \\
\hspace*{0.3cm} $qq\rightarrow WH$  &\POWHEGBOX[v2] & NLO & \PDFforLHC[15nlo] &  \Pythia[8.212] & AZNLO & NNLO(QCD)+NLO(EW) \\
\hspace*{0.3cm} $qq\rightarrow ZH$  &\POWHEGBOX[v2] & NLO & \PDFforLHC[15nlo] &  \Pythia[8.212] & AZNLO & NNLO(QCD)+NLO(EW)$^{\dagger}$ \\
\hspace*{0.3cm} $gg\rightarrow ZH$  &\POWHEGBOX[v2] & NLO & \PDFforLHC[15nlo] &  \Pythia[8.212] & AZNLO & NLO+NLL \\
\hspace*{0.3cm} $t\bar{t}H$ &\POWHEGBOX[v2] & NLO & \NNPDF[3.0nlo] & \Pythia[8.230] & A14 & NLO(QCD)+NLO(EW) \\
\bottomrule
\end{tabular}}}
\end{center}
\end{table}

%
%
%
%
%
%
%
%


%
%

\section{Event selection and categorisation}

\label{sec:esel}

Events are selected in three separate signal regions (SRs), which remain unchanged with respect to Ref.~\cite{HDBS-2018-40}.
The \hadhad signal region (SR) targets fully hadronic decay modes of the $\tau$-lepton pair, where the presence of two oppositely charged hadronically decaying $\tau$ leptons ($\tau_\mathrm{had}$) is determined by detector signatures compatible with the expected visible decay products (\thv).
Two \lephad SRs target events with a leptonic decay of a $\tau$ lepton ($\tau_{\mathrm{lep}}$) into an electron or a muon, and an oppositely charged \thv.
The electron and muon channels are considered together.
The decay channel with both $\tau$ leptons decaying leptonically is not studied in this paper but is instead covered in Ref.~\cite{HDBS-2019-02-journal}.
A control region (CR) is defined to constrain the background from $Z$ bosons produced in association with two jets initiated by $b$ or $c$ quarks (referred to as \ZHF in the following), and top-quark pair production processes.
In all regions the presence of two $b$-jets is also required.
These four regions are briefly summarised in Section~\ref{sub:presel}.
Selected events are split into different categories to enhance the sensitivity to the coupling modifiers \kl and \kvv, as described in Section~\ref{sub:cats}.
In each category a multivariate approach based on boosted decision trees (BDTs) is adopted to build the final discriminants, as detailed in Section~\ref{sub:bdt}.
The identification and reconstruction of electrons, muons, \thv, jets from the hadronisation of quarks and gluons, $b$-tagged jets, and missing transverse momentum ($\vec{p}_{\text{T}}^{\mathrm{miss}}$) is identical to what was documented in Ref.~\cite{HDBS-2018-40}.

\subsection{Preselection} %
\label{sub:presel}

Events in the \hadhad SR are selected using a combination of single-\thv triggers (STTs) and di-\thv triggers (DTTs), and are required to have two \thv with opposite charge.
An electron and muon veto is applied to ensure orthogonality with the \lephad SRs.
In the \hadhad SR event selection, the offline\footnote{In this paper, offline objects are those reconstructed after the data were collected, as opposed to trigger-level objects.} transverse momentum (\pT) thresholds for the \thv range between \qty{100}{\GeV} and \qty{180}{\GeV} for STT events depending on the data-taking period, while they are set at 40~GeV (30~GeV) for the (sub-)leading \thv for DTT events.
For events selected by the STTs a second \thv-candidate is required, with an offline \pT threshold of $25~\text{GeV}$.
Additional offline requirements for the DTTs are that either one extra jet with an offline \pT threshold set to 80~GeV is in the event, and the \thv are reconstructed within $\Delta R=2.5$ of each other, or that two extra jets with offline \pT thresholds set to \qty{45}{\GeV} are present in the event.
For events that satisfy both the STTs and DTTs, the offline requirements used for the STTs are applied.

Events containing exactly an electron or muon and one \thv with opposite charge are split into two mutually exclusive SRs, depending on whether they satisfy a single-lepton trigger (SLT) or a lepton-plus-\thv trigger (LTT), named the \lephad SLT SR and the \lephad LTT SR, respectively.
Only events failing the \lephad SLT SR selection are considered for the \lephad LTT SR\@.
Depending on the data-taking period, the offline electron (muon) selected by the SLT is required to have $p_{\text{T}}^{e} > 25$~GeV or $p_{\text{T}}^{e} > 27$~GeV ($p_{\text{T}}^{\mu} > 21$~GeV or $p_{\text{T}}^{\mu} > 27$~GeV).
Events selected in the \lephad LTT SR are required to contain either an electron or a muon with offline \pT thresholds set to $p_{\text{T}}^{e} > 18$ GeV and $p_{\text{T}}^{\mu} > 15$~GeV respectively, along with a \thv with an offline \pT threshold set to \qty{30}{\GeV}.

Events in all SRs are required to have $m^{\text{MMC}}_{\tau\tau} > 60$~GeV.\footnote{The invariant mass of the $\tau$-lepton pair ($m^{\text{MMC}}_{\tau\tau}$) is estimated by using the Missing Mass Calculator (MMC)~\cite{Elagin:2010aw}.}
To target $H\rightarrow \bb$ decays, events are required to contain exactly two $b$-tagged jets in the pseudorapidity region of $|\eta|<2.5$, satisfying the criteria of the `DL1r' $b$-tagging algorithm with a nominal efficiency of 77\% for $b$ jets~\cite{FTAG-2019-07}.
The two selected $b$-tagged jets have to satisfy minimum \pT thresholds of 45 and 20~GeV respectively, in addition to any trigger-dependent requirements.
In the \lephad SRs the invariant mass of the $b$-tagged jet pair ($m_{bb}$) is required to be lower than $150~\text{GeV}$ to reject background \ttbar events, and a \thv with \pT of at least 20~GeV is required in the \lephad SLT SR, while \pT $>$ 30~GeV is required in the \lephad LTT SR, in addition to any trigger-dependent requirements.

\begin{figure}[!tbh]
\centering
\includegraphics[width=0.5\textwidth]{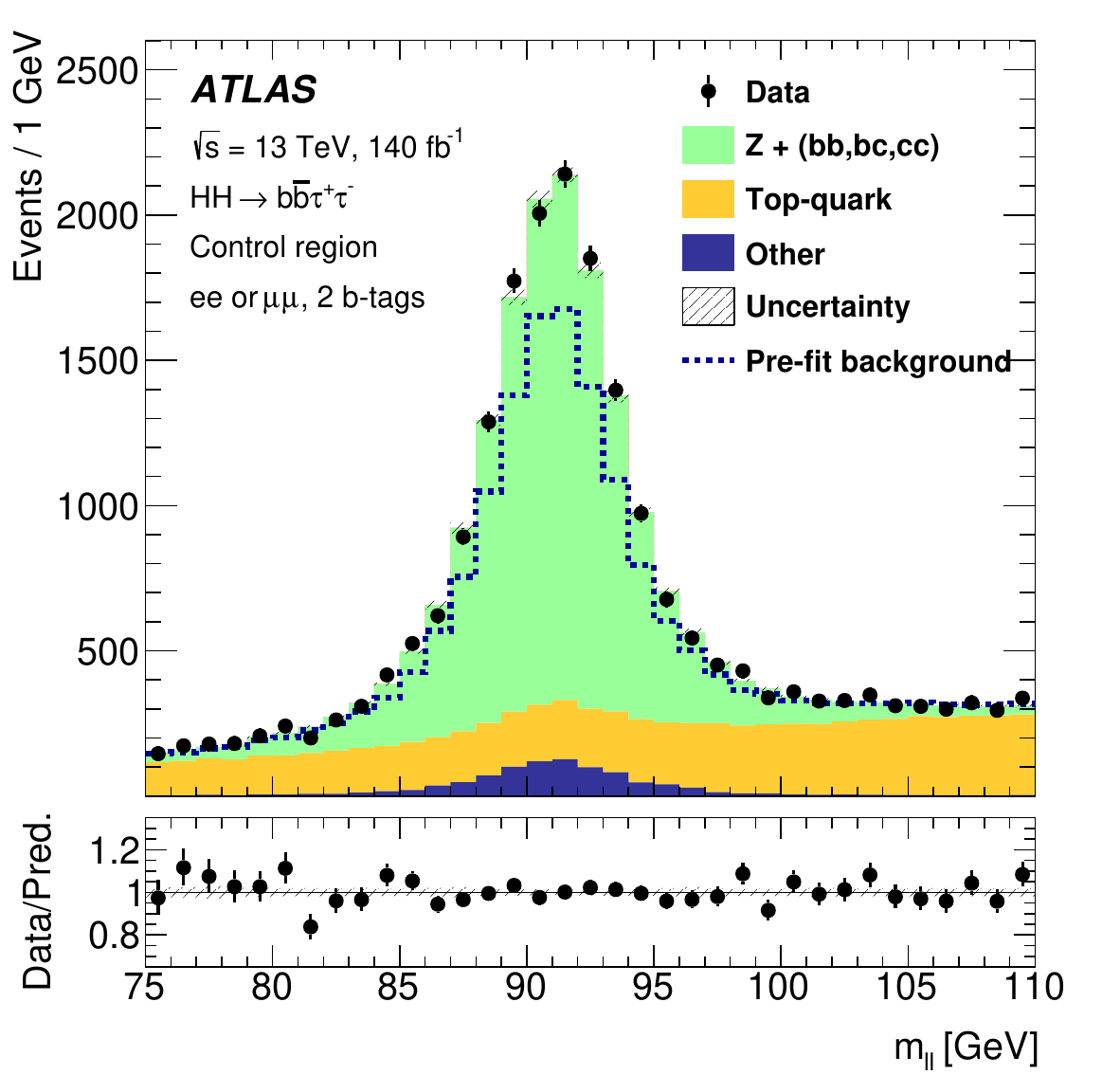}
\caption{Predicted and observed $m_{\ell\ell}$ distributions in the CR, with the normalisation, shape and total uncertainty of the backgrounds as determined from the likelihood fit to data described in Section~\ref{sec:result}.
The background processes named ``Other'' contain contributions from $Z$-boson production in association with less than two jets initiated by $b$ or $c$ quarks, $W$-boson production, vector boson pair production, $t\bar{t}W$ and $t\bar{t}Z$ production, and single Higgs boson production processes.
The dashed histogram shows the total pre-fit background.
The lower panel shows the ratio of data to the total post-fit background.
The hatched bands in the upper and lower panels show the combined statistical and systematic uncertainties in the total background.
}
\label{fig:mllCR}
\end{figure}

Events in the CR are required to contain exactly two electrons or two muons of opposite charge with a dilepton invariant mass ($m_{\ell\ell}$) within the range of $75~\text{GeV} < m_{\ell\ell} < 110~\text{GeV}$, and exactly two $b$-tagged jets.
The $m_{bb}$ is required to be less than $40~\text{GeV}$ or greater than $210~\text{GeV}$ to avoid overlap with other analyses targeting $H\to b\bar{b}$ decays.
Compared with Ref.~\cite{HDBS-2018-40}, the requirement on the transverse momentum of the selected leptons is raised to $\pT > 40~\text{GeV}$ (from $\pT > 9~\text{GeV}$), while the leading $b$-tagged jet is required to have $\pT > 45~\text{GeV}$.
This selection provides a closer alignment between the kinematic properties of events selected in the CR and the SRs. Figure~\ref{fig:mllCR} shows the predicted and observed $m_{\ell\ell}$ distributions in the CR, after the likelihood fit described in Section~\ref{sec:result}.

The main sources of background in the SRs after this preselection are from top-quark, $Z+$jets, $W+$jets, diboson, single Higgs boson and multi-jet production.
Depending on the source, the background contamination is estimated by using data-driven or simulation-based techniques, or a combination of both.
The normalisations of simulated \ttbar  and \ZHF backgrounds are determined from data in the likelihood fits of signal and control regions described in Section~\ref{sec:result}.
A reconstructed \thv candidate in these background events can originate either from a \thad decay (true-\thv), or from a misidentified quark- or gluon-initiated jet (fake-\thv).
Events in which an electron or a muon is misidentified as a \thv represent a small additional background.
The processes that contribute most to background events with fake-\thv candidate are \ttbar and multi-jet production.
In \ttbar  events, fake-\thv candidates typically originate from quark-initiated jets from the top-quark decay.
In multi-jet events, both quark- and gluon-initiated jets are a source for fake-\thv candidates.
Events with fake-\thv candidate in \ttbar  and multi-jet production are estimated from techniques relying on both simulated events and data, described in Ref.~\cite{HDBS-2018-40}, which prove to provide an accurate modelling of the variables used for the event categorisation and multivariate techniques described in this Section.
The estimate of this background relies on a good description of the fundamental properties of $\tau$ leptons in the three analysis SRs, which are unchanged compared to Ref.~\cite{HDBS-2018-40}.
In the \lephad regions a combined fake-factor method is used to estimate \ttbar and multi-jet events with fake-\thv, by employing two groups of regions:
the identification (ID) regions containing one identified \thv, and the anti-identification (anti-ID) regions containing one reconstructed \thv with modified requirements (anti-\thv), leading to an enriched fake-\thv contribution.
Fake-factors are derived separately for \ttbar and multi-jet events in dedicated ID and anti-ID control regions, and are combined to scale events from the anti-ID SRs to obtain the fake-\thv background prediction in the \lephad SRs.
In the \hadhad regions two separate methods are used to estimate the backgrounds with fake-\thv from \ttbar and multi-jet production.
Multi-jet events can only enter the SRs when both \thv are fake, and their contribution is estimated by using a fake-factor method.
Fake-factors are derived for the multi-jet background in a dedicated set of ID control regions, defined for events with two identified \thv, and anti-ID control regions for events with one identified \thv and one anti-\thv candidate.
The fake-factors are applied to scale events from the anti-ID SRs to obtain the multi-jet fake-\thv background prediction in the \hadhad SRs.
Background events with fake-\thv from \ttbar production in the \hadhad SRs are estimated using simulation, while correcting the fake-\thv misidentification efficiencies with scale factors derived from data in the \ttbar control region defined for the fake-factor estimate in the \lephad channel.
The modelling of events with a fake-\thv candidate from background processes other than \ttbar and multi-jet production is performed using MC simulation, as they represent a minor contribution to the total background.
Simulated event samples are used to model background events containing true-\thv and events with an electron or a muon misidentified as a \thv candidate.
The changes introduced to the MC simulation detailed in Section~\ref{sec:mc} are found to have a negligible impact on the data-driven estimate of the background.

\subsection{Event categorisation}
\label{sub:cats}

\begin{figure}[!tbh]
\centering
\includegraphics[width=\textwidth]{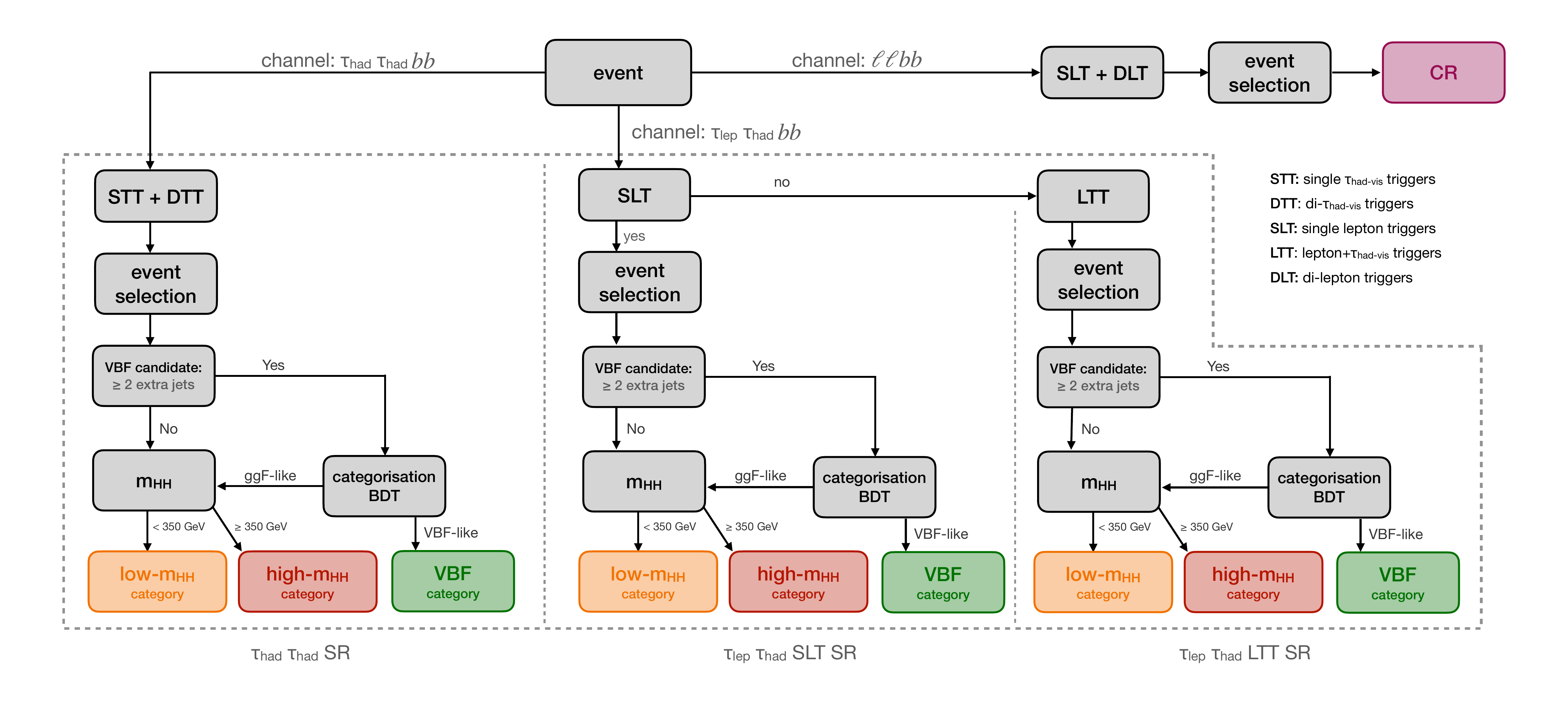}
\caption{Flowchart summarizing the definition of the \hadhad SR, the \lephad SLT SR, the \lephad LTT SR and the dedicated CR defined in Section~\ref{sec:esel}.
The flowchart shows the selection criteria applied to define the corresponding ggF low-\mhh, ggF high-\mhh and VBF categories for each analysis SR, leading to a total number of nine analysis categories and an additional CR.
}
\label{fig:SRscheme}
\end{figure}

Events selected in each SR described in Section~\ref{sub:presel} are split into three separate categories.
To enhance the sensitivity to the coupling modifier \kvv, a dedicated VBF category is defined with a multivariate approach, defining a dedicated BDT to select events with characteristic features of VBF \HH production, separately for each analysis SR\@.
These are referred to as \textit{categorisation BDTs} in the following.
The distribution of the invariant mass of the \HH system (\mhh) in ggF \HH events is significantly affected by the value of \kl.
Hence, events not falling in the VBF category are split into two \mhh categories, targeting ggF \HH production with \kl values close to the SM expectation (ggF high-\mhh) or significantly different from it (ggF low-\mhh).
The three categories are mutually exclusive, and they are defined separately for each SR following the procedure outlined in Figure~\ref{fig:SRscheme}, leading to a total of nine event categories.
First, VBF candidate events are identified by requiring the presence of at least two jets in addition to the ones associated with the $H\to b\bar{b}$ decay.
These events are used to train the categorisation BDTs to separate ggF and VBF \HH production modes in event topologies with additional jets.
Events satisfying a SR-dependent selection on the output of these BDTs are assigned to the VBF category.
Events failing the VBF selection are assigned to the ggF categories, together with events not selected as VBF candidates.

The categorisation BDTs are built using the Toolkit for Multivariate Data Analysis (TMVA~\cite{TMVA}) using ggF \HH events as signal and VBF \HH events as background, scaled such that the total event yield is the same for both processes.
A dedicated version of the BDT is trained in each SR respectively.
To make use of the complete set of simulated events for the BDT training, optimisation and evaluation, the events are split into three samples of equal size, A, B and C.
The performance of the BDTs trained on sample A (respectively B, C), is optimised using sample B (C, A) and evaluated with sample C (A, B), ensuring an unbiased estimate of the performance of the BDTs.
The selected data events are split into three samples, each analysed with one of the three separately trained BDTs.
The output distributions of the BDTs evaluated on samples A, B and C are merged for both simulated and data events to produce the final discriminant.
The number of trees and their depth are chosen to maximise the BDT separation power, quantified by the value of the number-counting significance $z$~\cite{Cowan:2010js} computed from the binned distribution of the BDT discriminant.

\begin{figure}[!tbh]
\centering
\subfigure[]{\includegraphics[width=0.32\textwidth]{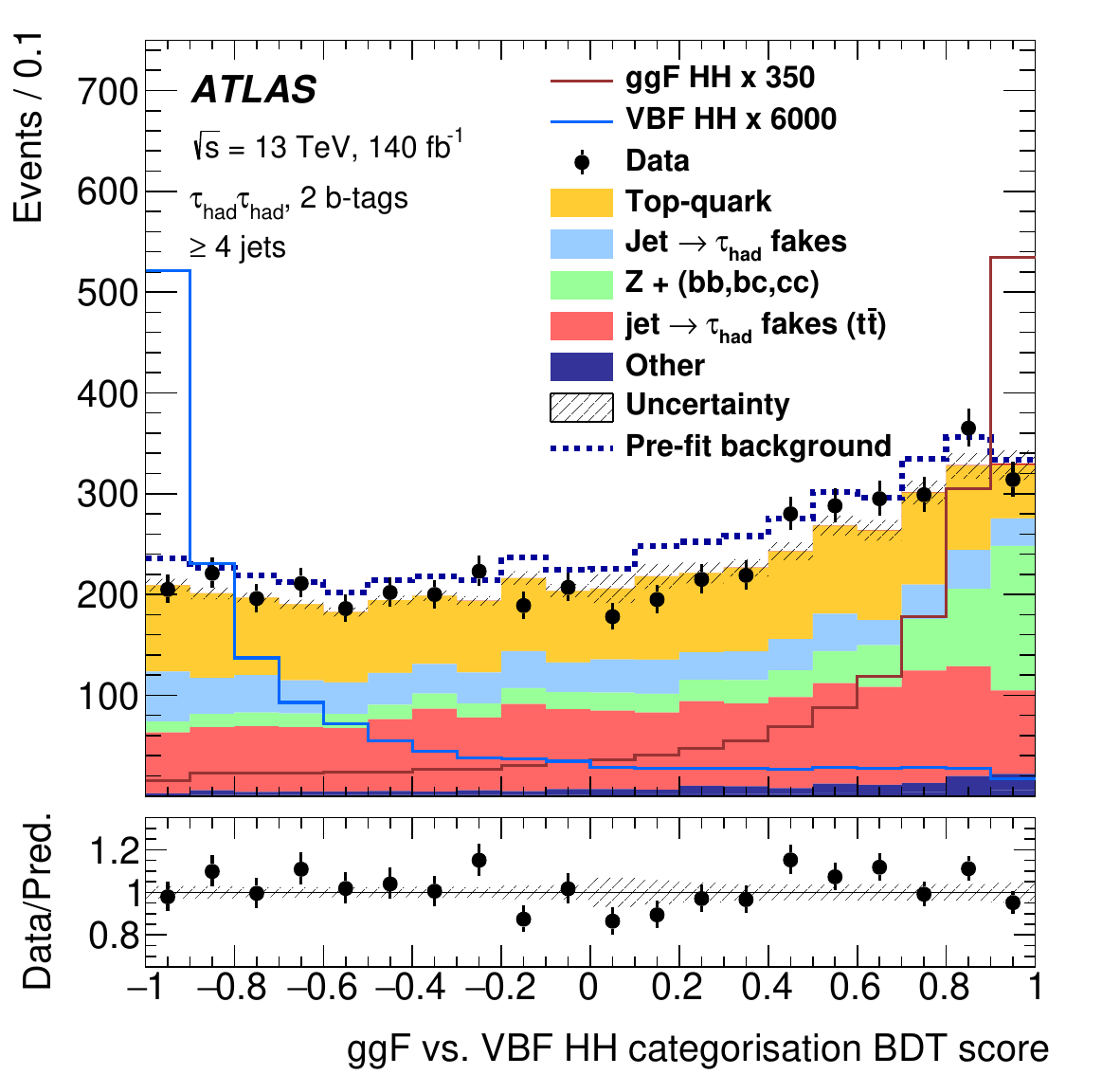}}
\subfigure[]{\includegraphics[width=0.32\textwidth]{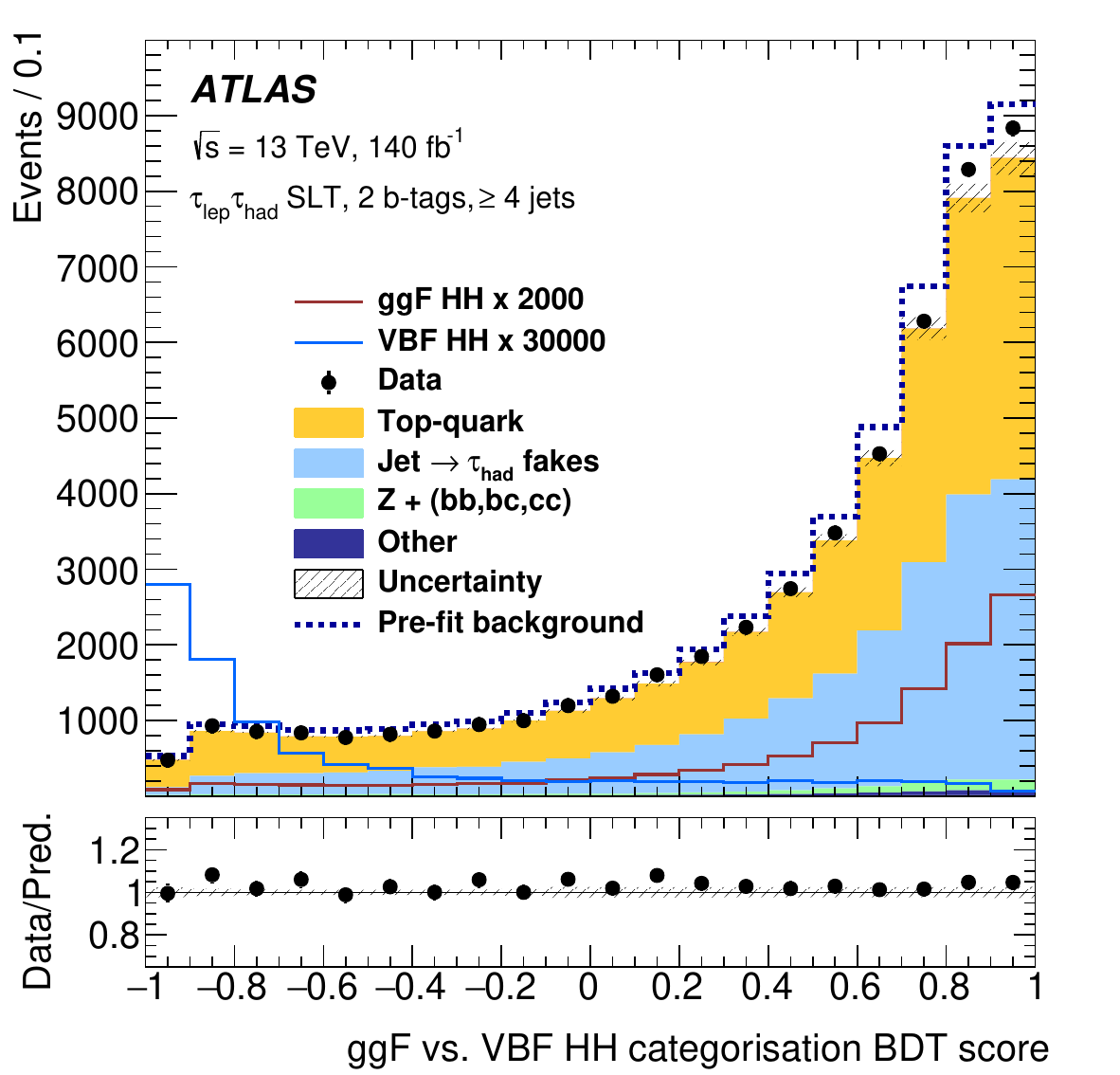}}
\subfigure[]{\includegraphics[width=0.32\textwidth]{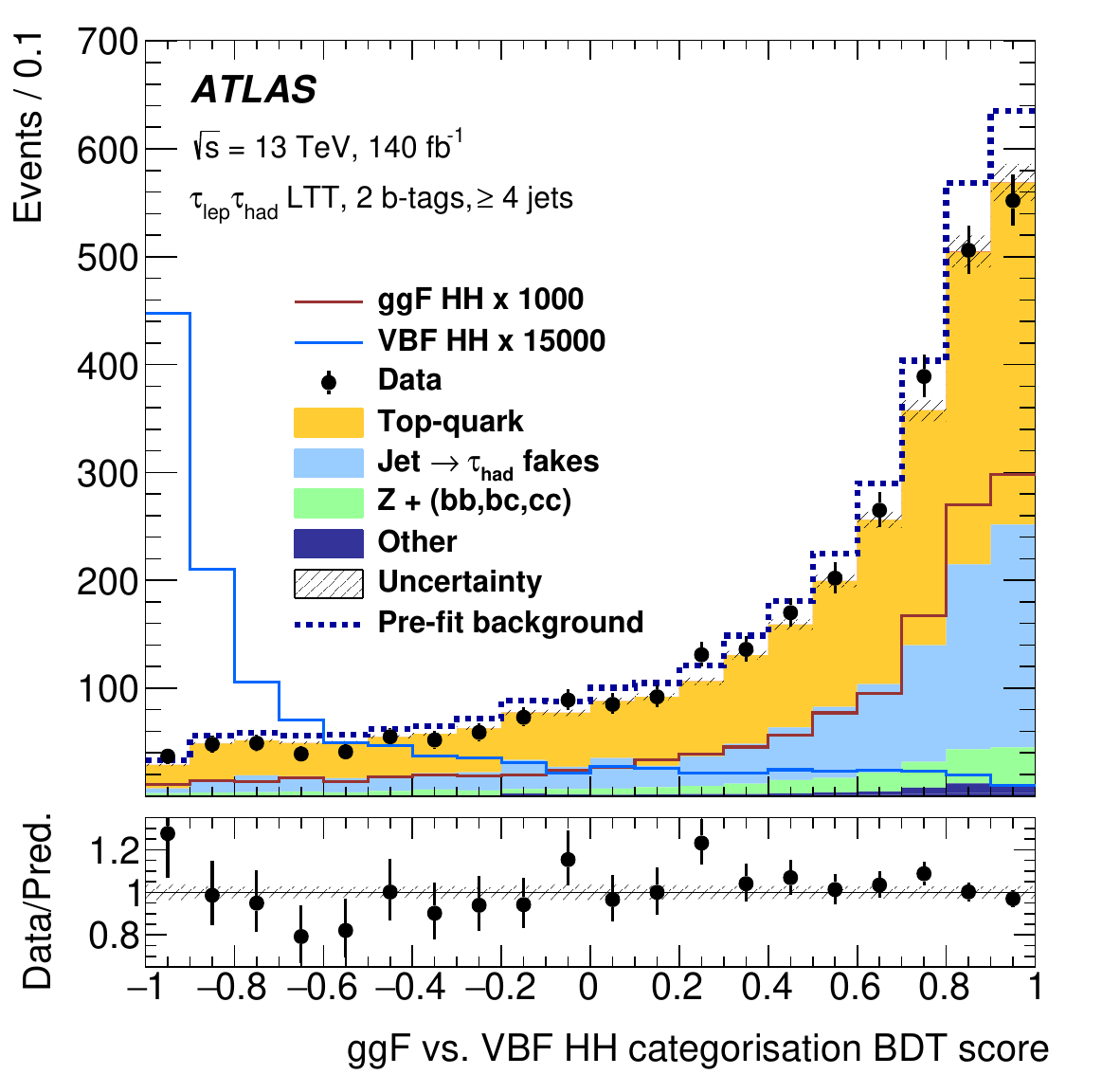}}
\caption{Predicted and observed distributions of the categorisation BDT scores in~(a) the \hadhad, (b) the \lephad SLT and~(c) the \lephad LTT SRs, for all candidate VBF events.
The background distributions are shown as obtained from the combined likelihood fit to data described in Section~\ref{sec:result}.
The background processes named ``Other'' contain contributions from $Z$-boson production in association with less than two jets initiated by $b$ or $c$ quarks, $W$-boson production, vector boson pair production, $t\bar{t}W$ and $t\bar{t}Z$ production, and single Higgs boson production processes.
The ggF and VBF \HH signal distributions are overlaid and scaled to the factor indicated in the legend times the SM expectation.
The dashed histogram shows the total pre-fit background.
The lower panels show the ratio of data to the total post-fit sum of signal and background.
The hatched bands in the upper and lower panels show the combined statistical and systematic uncertainties in the total prediction.
}
\label{fig:catBDTs}
\end{figure}

The minimal set of training variables that optimises the BDT separation power is determined starting from a small set and iteratively adding variables one at a time from a pre-defined list of candidate variables.
The variable leading to the largest increase (or minimal decrease\footnote{Variables showing minimal decrease of the $z$ significance are retained to mitigate the impact of statistical fluctuations on the optimisation, potentially leading to prematurely terminating the iterative procedure. In the final selection only variables improving the BDT sensitivity are retained.}) in significance $z$ is included, until no changes are observed.
The starting set of variables comprises the invariant mass of the VBF jets ($\mjj^{\mathrm{VBF}}$), defined as the two jets with the highest \pT not associated with the $H\to b\bar{b}$ decay, and their pseudorapidity gap ($\dEtajj^{\mathrm{VBF}}$).
The final set of variables for the categorisation BDTs in each SR is summarised in Table~\ref{tab:catBDTvars}.
It includes the product of VBF jet pseudorapidities~(VBF $\eta_0 \times \eta_1$), their angular separations ($\dPhijj^{\mathrm{VBF}}$ and  $\dRjj^{\mathrm{VBF}}$) and \mhh.
In addition, the Fox-Wolfram moments $f_i$ of $i$-th order~\cite{Bernaciak_2013} and their modified definitions for usage in hadron collider experiments $h_i$~\cite{spill:2016} can further increase the separation power, together with the centrality\footnote{The centrality of a set of four-momenta of index $i$ is defined as $C = \frac{\sum\limits_i \pT(i)}{\sum\limits_i E(i)}$} ($C$) and the invariant mass ($m_{\mathrm{Eff}}$) of the system composed by the selected $\tau$ leptons, the missing transverse momentum vector $\vec{p}_{\text{T}}^{\mathrm{miss}}$ and the selected jets.
The predicted and observed distributions of the resulting BDT scores are shown in Figure~\ref{fig:catBDTs} for all three SRs.
VBF candidate events are assigned to the VBF category if their BDT score is evaluated below a certain threshold.
The value of this threshold is optimised to achieve the best upper limits on \HH production for ggF and VBF production modes separately and combined, along with the best exclusion limits for the coupling modifiers \kl and \kvv from the likelihood fit described in Section~\ref{sec:result}.
The categorisation BDT cut values are set to $0.1$, $-0.13$ and $-0.1$ for the \hadhad, \lephad SLT and \lephad LTT SRs, respectively.

\begin{table}[!tbh]
\centering
\caption{Input variables for the categorisation BDTs in each of the three SRs.
The superscripts $a$ and $c$ specify the selection of jets that are taken into account for the calculation in addition to the two $\tau$-lepton candidates and the $\vec{p}_{\text{T}}^{\mathrm{miss}}$ vector.
For variables with a $c$, only the four-momenta of central jets, i.e., jets with $|\eta| < 2.5$, are included, while an $a$ indicates that all available jets are included.
}
\small
\label{tab:catBDTvars}
\renewcommand{\arraystretch}{1.5}%
\begin{tabular}{lccc}
\toprule
Variable & \hadhad & \lephad SLT & \lephad LTT\\
\midrule
$\mjj^{\mathrm{VBF}}$ & \ding{51} & \ding{51} & \ding{51}\\%
$\dEtajj^{\mathrm{VBF}}$ & \ding{51} & \ding{51} & \ding{51}\\%
VBF $\eta_0 \times \eta_1$ & \ding{51} & \ding{51} &\\%
$\dPhijj^{\mathrm{VBF}}$ & \ding{51} & &\\%
$\dRjj^{\mathrm{VBF}}$ & & \ding{51} & \ding{51}\\%
\dRtautau & \ding{51} & &\\%
\mhh & \ding{51} & &\\
$f_2^a$ & \ding{51} & &\\%
$C^a$ & & \ding{51} & \ding{51}\\ %
$m_{\mathrm{Eff}}^a$ & & \ding{51} & \ding{51}\\
$f_0^c$ & & \ding{51} & \\%
$f_0^a$ & & & \ding{51}\\%
$h_3^a$ & & & \ding{51}\\%
\bottomrule
\end{tabular}
\end{table}

Events not retained in the VBF category are split into low-\mhh and high-\mhh categories targeting ggF \HH production.
While the ggF \HH cross-section increases for \kl values larger than the SM expectation and close to the current experimental sensitivity ($\kl \sim 6$~\cite{HDBS-2022-03}), the softer \mhh spectrum leads to smaller detector acceptance and selection efficiency.
The event split into different regions of \mhh allows to partially disentangle these effects, improving the sensitivity to higher \kl values.
The splitting value is chosen to be \qty{350}{\GeV} since the effect of the interference between the box and triangle diagrams on the differential ggF \HH production cross-section is maximal there.
\subsection{Multivariate discriminants} %
\label{sub:bdt}
Based on the event categorisation described in Section~\ref{sub:cats}, an additional set of multivariate discriminants making use of BDTs is trained and evaluated in each of the analysis SRs separately to achieve optimal separation between the \HH signal and the background.
One dedicated BDT is trained for each analysis SR (\hadhad, \lephad SLT, \lephad LTT) and each category (ggF low-\mhh, ggF high-\mhh and VBF), leading to nine separate BDTs.
In the ggF high-\mhh and VBF categories, the signal corresponds to the ggF and VBF SM production of \HH pairs, respectively.
In the ggF low-\mhh category the signal is defined as ggF \HH production with $\kl=10$.
During training, the sum of all backgrounds normalised to their respective cross-sections is used, which is scaled such that the total background yield matches that of the corresponding signal sample.
The backgrounds containing \thv from misidentified quark- or gluon-initiated jets are modelled using simulation, except for the multi-jet background in the \hadhad category, where the data-driven estimate introduced in Section~\ref{sub:presel} is used.
The BDTs evaluated in the low-\mhh and high-\mhh categories are trained on the event samples selected in the respective categories.
The BDTs evaluated in the VBF categories are trained on events from both the VBF and ggF categories to maximise the available sample size.

To make use of the complete set of simulated events for the BDT training, optimisation and evaluation, the events are split into three samples, following the same procedure described in Section~\ref{sub:cats}.
The number of trees and their depth are chosen to maximise the BDT separation power, quantified by the value of the signal significance $z$ computed from the binned distribution of the BDT output.

The input variables used for the BDTs are chosen to maximise the BDT separation power, separately for each trained BDT\@.
A list of variables is built for each BDT, ordered according to their impact on the signal significance $z$, following the same procedure as described in Section~\ref{sub:cats}.
In each analysis SR and category the starting set comprises the following variables: the invariant mass of the two selected $b$-jets (\mbb), the invariant mass of the $\tau$-lepton pair (\mMMC), the \mhh reconstructed from the selected $b$-jet and $\tau$-lepton pairs, the angular separation between $b$-jets (\dRbb) and between $\tau$-leptons (\dRtautau).
In the \lephad LTT high-\mhh category \dRbb is not included, and in the \lephad LTT VBF category both \dRbb and \dRtautau are removed.
The additional variables included in the BDT training through a $z$-based optimisation as described in Section~\ref{sub:cats} can be grouped in several categories. Variables that require the presence of a charged lepton are not included for the \hadhad SR\@.
Higgs boson candidates $H$ are reconstructed from either $b$-jet or $\tau$-lepton pairs.
For the low-\mhh and high-\mhh categories, variables describing the kinematic properties of the selected $b$-jets and $\tau$ leptons are included: the transverse momentum of the leading and subleading $b$-jets and $\tau$ leptons, along with the pseudorapidity of the $\tau$ leptons and the transverse mass of each of the $\tau$ leptons and $\vec{p}_{\text{T}}^{\mathrm{miss}}$.
The angular separation between the (sub)leading $b$-jet and the (sub)leading $\tau$ lepton is included, along with the angular separation between the leading $b$-jet and the subleading $\tau$ lepton, the pseudorapidity separation and the transverse momentum difference between the selected $\tau$-lepton candidate and the charged lepton.
Variables related to the reconstructed $H$ candidate topologies include: the azimuthal angular separation between the $b$-jet pair and the $\tau$-lepton pair, and between either the $b$-jet or the $\tau$-lepton pair and the magnitude $E_{\text{T}}^{\mathrm{miss}}$ of $\vec{p}_{\text{T}}^{\mathrm{miss}}$, the azimuthal angle of the selected $b$-jet (or $\tau$-lepton) pair relative to the $H$ candidate rest frame, along with the transverse momentum of the $b$-jet and $\tau$-lepton pairs.
The separation power is further increased by including variables describing the hadronic activity in candidate events, such as the number of selected hadronic jets, the scalar sum of jet transverse momenta ($H_{\text{T}}$), the effective mass of the $\tau$-lepton pair and all jets with $|\eta|<2.5$, the $E_{\text{T}}^{\mathrm{miss}}$, the stransverse mass~\cite{Lester_2011} ($M_{T2}$), the azimuthal angular separation between the selected charged lepton and the $\vec{p}_{\text{T}}^{\mathrm{miss}}$.
Additional variables are included to characterise event properties, such as the transverse mass of the $W$ boson candidate in the \lephad SRs (defined as the transverse mass of the selected lepton and the $\vec{p}_{\text{T}}^{\mathrm{miss}}$), the topness variable~\cite{PhysRevLett.111.121802} as defined in Ref.~\cite{Kim_2019} assuming either $\sigma_t = \qty{5}{\GeV}, \sigma_W = \qty{5}{\GeV}$ ($T_1$) or $\sigma_t = \qty{15}{\GeV}, \sigma_W = \qty{5}{\GeV}$ ($T_2$), the reduced invariant mass of the \HH system (defined as $\mhh^\ast = \mhh - \mbb - \mMMC + 250$~GeV), the scaled invariant mass of the \HH system (defined by scaling the four-momenta of both $H$ candidates by the ratio $125~$GeV$/m_H$, where $m_H$ is the $H$ candidate invariant mass), the transverse momentum of the reconstructed \HH system and the effective mass of the \HH decay products.
To construct the $E_{\text{T}}^{\mathrm{miss}}$ centrality, the transverse plane is transformed such that the directions of the \thv candidates or leptons are orthogonal, and that the smaller $\phi$ angle between them defines the positive quadrant of the transformed plane.
Then the $E_{\text{T}}^{\mathrm{miss}}$ centrality is defined as the sum of the $x$ and $y$ components of the $\vec{p}_{\text{T}}^{\mathrm{miss}}$ unit vector in this transformed plane.
Dedicated variables describing characteristic event configurations are included via a selection of Fox-Wolfram moments (as defined in Section~\ref{sub:cats}), circularity, sphericity, thrust and planar flow variables reconstructed from the \HH decay products.
Finally, the $b$-tagging information is provided by the quantile distribution of the DL1r tagger output for the selected $b$-jets, which is included as a training variable.
For the VBF categories the first five variables listed in Table~\ref{tab:catBDTvars} are also included, along with circularity, sphericity and planar flow variables reconstructed from the $\tau$-lepton pair and all selected jets, to target the specific features of VBF \HH events.

%
%
%
%
%
%

%
%
%


%
%

\section{Systematic uncertainties}

\label{sec:systs}
While the largest source of uncertainty in this analysis comes from the limited number of data events, systematic uncertainties can affect the signal and background estimates. %
Uncertainties originating from the detector response in the selection and reconstruction of the used objects are included as documented in Ref.~\cite{HDBS-2018-40}.
Statistical uncertainties in the predicted background processes are modelled using a simplified version of the Beeston-Barlow technique~\cite{Barlow:1993dm}, in which only the uncertainty in the total background contribution in each bin is considered.
An uncertainty in the full Run 2 integrated luminosity of 0.83\%~\cite{DAPR-2021-01}, from measurements using the LUCID-2~\cite{LUCID2} detector and a complementary measurement using the inner detector and calorimeters, is assigned to physics processes whose normalisations are taken from simulation.

For all processes whose estimation relies on MC simulation, the impact of various sources of theoretical uncertainties in their cross-section, on the fractional contribution to each analysis category within each SR, as well as on the shape of the BDTs introduced in Section~\ref{sub:bdt} is considered.
The total normalisation of \ttbar and \ZHF backgrounds is determined via the likelihood fit described in Section~\ref{sec:result}; therefore, no uncertainty is included for their cross-section calculation.
For \ttbar and \ZHF backgrounds, the uncertainty in the fractional contribution in each analysis category is computed as the relative variation in acceptance with respect to the dedicated CR introduced in Section~\ref{sec:esel}, while for other processes this is evaluated as the absolute acceptance variation in each analysis category.
To assess an uncertainty in the shape of the BDT output, a dedicated rebinning algorithm is applied to ensure that only statistically significant shape variations are considered.
In the signal-enriched region of the BDT distribution, the statistical uncertainty of each bin is required to be below a process- and variation-dependent threshold (ranging from 5\% to 15\%), while in the background-enriched region of the distribution the fraction of total events per bin is required to be larger than 5\%.
For each process, the impact of each source of uncertainty in the fractional contribution in each analysis category is correlated with the uncertainty in the shape of the BDT score in the corresponding category, in the likelihood fit described in Section~\ref{sec:result}.
All sources of uncertainty are evaluated separately and correlated across the three analysis SRs.
The uncertainties in the SM cross-section calculations for all processes are unchanged with respect to Ref.~\cite{HDBS-2018-40}, along with the full uncertainty estimate for smaller background processes including $Z$+light-flavour jets, $W$+jets, diboson and single-top-quark production in the $s-$ and $t-$channels.

The \ttbar hard-scatter and parton-shower uncertainties are evaluated by comparing the nominal sample with samples generated using \MGNLO+\Pythia[8] and \POWHEGBOX[v2]+\HERWIG[7.04]~\cite{Bahr:2008pv,Bellm:2015jjp}, respectively.
The hard-scatter uncertainty also accounts for uncertainties in the matching and merging of the matrix-element calculation with the parton-shower algorithm.
The uncertainty in missing higher-order QCD corrections and the modelling of initial-state QCD radiation is assessed via independent variations of the renormalisation and factorisation scales in the hard-scatter calculation, of the showering tune \textsc{Var3c} parameter~\cite{ATL-PHYS-PUB-2014-021} and of the $h_{\mathrm{damp}}$\footnote{The $h_{\mathrm{damp}}$ parameter regulates the $p_{\text{T}}$ of the first additional emission beyond the Born configuration in the \POWHEGBOX generator, controlling the matching of the matrix element to the parton shower.} parameter, while for the modelling of final state QCD radiation alternative choices of factorisation and renormalisation scales in the showering algorithm are considered.
Finally, uncertainties in the PDF and the value of the strong coupling constant $\alpha_S$ are also evaluated.
All sources of uncertainty have an impact on the shape of the BDT score and the \ttbar fractional contribution in each analysis category.
The largest effect on the latter is due to parton-shower variations and ranges between 1\% and 10\% of the nominal values, depending on the analysis category and SR.

For \ZHF processes the uncertainty in the modelling of the hard scatter and the parton shower are evaluated by comparing the nominal sample with a \MGNLO+\PYTHIA8 sample with up to three additional partons in the final state at NLO accuracy in the QCD coupling, in which additional jet multiplicities are merged together with the FxFx NLO matrix-element and parton-shower merging prescription~\cite{Frederix:2012ps}.
The A14 parton-shower tune and the \NNPDF[2.3lo] PDF set are used for this alternative sample.
Uncertainties from missing higher-order QCD corrections are evaluated with renormalisation and factorisation scale variations from the nominal samples, along with PDF and $\alpha_S$ variations.
The effect of higher-order electroweak corrections for \ZHF processes is found to be negligible, and thus not included.
Uncertainties in the matching between matrix element calculation and parton shower are considered via variations  of the \SHERPA matching parameter (CKKW) and the resummation scale (QSF).
All sources of uncertainty have an impact on the \ZHF fractional contribution in each analysis category, with the largest effect due to renormalisation and factorisation scale variations ranging up to 13\% of the nominal values depending on the analysis category and SR\@.
The uncertainty in the modelling of the hard scatter and the parton shower, from the comparison with the alternative \MGNLO+\PYTHIA8, is the only source of uncertainty with a significant impact on the shape of the BDT score in the analysis SRs.
The same source of uncertainty is found to have a non-negligible impact on the shape of the $m_{\ell\ell}$ variable in the CR, which is included as a dedicated uncertainty uncorrelated from the BDT score shape uncertainty in the likelihood fit.
Finally, an additional systematic uncertainty is included to cover the residual difference between data and MC simulation in the dedicated CR, as a function of the transverse momentum of the selected lepton pair.
This uncertainty is applied in all analysis SRs as a function of the transverse momentum of the $\tau$-lepton pair selected from the MC truth record.

For single-top-quark processes only uncertainties related to the $Wt-$channel are considered, as it is found to be dominant compared with $s-$ and $t-$ channel contributions.
The hard-scatter and the parton-shower uncertainties are evaluated by comparing the nominal sample with alternative \MGNLOPY{8} and \POWBOXHER{7} samples.
The uncertainty in the modelling of QCD radiation is evaluated by varying the showering tune \textsc{Var3c}, along with independent variations of the renormalisation and factorisation scales in the hard-scatter and parton-shower calculations.
The uncertainty related to the interference between the \ttbar and the $Wt-$channel single-top-quark processes is evaluated by comparing the nominal $Wt$-channel sample produced with the diagram removal scheme to an alternative sample produced with the diagram subtraction scheme~\cite{Frixione:2008yi}.
Finally, uncertainties in the PDF are also evaluated.
All sources of uncertainty have an impact on the $Wt-$channel single-top-quark fractional contribution in each analysis category.
Variations due to the uncertainty in the $Wt-$channel interference scheme range from 1\% to 7\% in the low-\mhh categories, from 23\% to 29\% in the high-\mhh categories and from 14\% to 34\% in VBF categories.
Uncertainties in the $Wt-$channel interference scheme are also evaluated on the shape of the BDT score, rather than on the \pT of the $b$-tagged jet pair as in Ref.~\cite{HDBS-2018-40}.
The hard-scatter and parton-shower uncertainties range from 15\% to 36\% on the $Wt-$channel fractional contribution, depending on the analysis category and SR\@.
Uncertainties in the modelling of QCD radiation are found to have a significant impact on the shape of the BDT score.

An uncertainty of 100\% is applied on the normalisation of single Higgs boson in the ggF, VBF and associated production $WH$ mechanisms where the Higgs boson decays into a $\tau$-lepton pair, to account for difficulties in the modelling of these processes in association with heavy-flavour jets~\cite{TOPQ-2017-12,STDM-2012-11}.
Uncertainties from missing higher-order QCD corrections are evaluated with independent variations of the renormalisation and factorisation scales from the nominal samples, along with PDF and $\alpha_S$ variations.
Parton-shower uncertainties are evaluated by comparing the nominal sample to alternative \POWBOXHER{7} samples for associated production $ZH$ and $\ttbar H$ processes.
For $\ttbar H$ processes the hard-scatter uncertainties are derived by comparing the nominal samples to \MGNLOPY{8} predictions, and uncertainties in the modelling of QCD radiation are assessed via variations of the showering tune \textsc{Var3c}, along with independent variations of the renormalisation and factorisation scales in the parton-shower algorithm.
All sources of uncertainty have an impact on the single-Higgs-boson fractional contribution in each analysis category, while only parton-shower uncertainties are included as variations of the BDT score for the $ZH$ and $\ttbar H$ processes.

For the SM ggF and VBF \HH signal processes, uncertainties from missing higher-order QCD corrections are assessed via independent variations of the renormalisation and factorisation scales in the hard-scatter calculation, along with PDF and $\alpha_S$ variations.
Parton-shower uncertainties are evaluated by comparing the nominal samples with alternative \POWBOXHER{7} samples.
All sources of uncertainty have an impact on the signal fractional contribution in each analysis category, while only parton-shower uncertainties are included as variations of the BDT score. %
Cross-section uncertainties for single-Higgs-boson and \HH processes~\cite{Baglio:2020wgt} follow the same approach detailed in Ref.~\cite{HDBS-2018-40}, along with uncertainties in the $H\rightarrow\bbbar$ and $H\rightarrow\tau^+\tau^-$ branching ratios~\cite{deFlorian:2016spz}.

A dedicated uncertainty in the reweighting method applied to the ggF \HH samples to model alternative \kl hypotheses, described in Section~\ref{sec:mc}, is defined via a comparison of the SM ggF \HH samples with an alternative sample generated assuming $\kl = 10$.
The SM sample and the $\kl = 10$ sample are reweighted to a wide range of \kl values and the acceptance values of both predictions are compared independently in each category.
The maximum of the obtained deviations is taken as an uncertainty and applied to the $\kl = 0$ and $\kl = 20$ templates used for the linear combination of signal samples in the fit.
These deviations range from 2\% to 4\% in the ggF categories and from 20\% to 40\% in the VBF categories.
The uncertainty in the VBF categories arises from the \mhh-based reweighting method not accounting for additional radiation produced with the \HH pair, to which the categorisation BDTs are highly sensitive.
The linear combination of VBF \HH samples described in Section~\ref{sec:mc} is found to accurately model alternative \kl and \kvv hypothesis; therefore, no dedicated uncertainty is considered.

The estimate of systematic uncertainties affecting the data-driven background follows the approach described in Ref.~\cite{HDBS-2018-40}.
Uncertainties in the fractional contribution of the data-driven background and on the shape of the BDT score are derived in all the analysis SRs and categories.


%
%

\section{Results}

\label{sec:result}

The statistical procedures applied for extracting results are the same as in Refs.~\cite{HDBS-2018-40, HDBS-2022-03}.
The global likelihood function $L(\alpha, \theta)$ is constructed from the binned distribution of the BDT output score for signal, background and data distributions in the nine orthogonal analysis categories described in Section~\ref{sec:esel}, together with the \mll distribution in the dedicated CR.
The set $\alpha$ contains the parameters of interest (POI) of the measurement, while $\theta$ is the ensemble of nuisance parameters, corresponding to systematic uncertainties constrained by auxiliary measurements in control regions or by theoretical predictions, or to parameters such as the $\ttbar$ and \ZHF background yields, which are \textit{a priori} unconstrained.

A dedicated procedure is applied to transform the BDT discriminant to obtain a smoother distribution for the background processes and a finer binning in the regions with the largest signal contribution, while preserving a sufficiently large number of background events in each bin, similarly to Ref.~\cite{HDBS-2018-40}.
Starting from finely binned histograms, bins are iteratively merged starting from the most signal-enriched bins until they satisfy the condition of $10 f_{s} + 5 f_{b} > 1$, where $f_{s}$ and $f_{b}$ are the fractions of signal and background contained in the bin, respectively.
The relative statistical uncertainty in the predicted background contribution has to be below 20\%, and at least three expected background events are required per bin.
The \mll distribution in the CR is binned uniformly with a width of \qty{1}{\GeV}.

\begin{figure}[!tbh]
\centering
\subfigure[]{\includegraphics[width=0.32\linewidth]{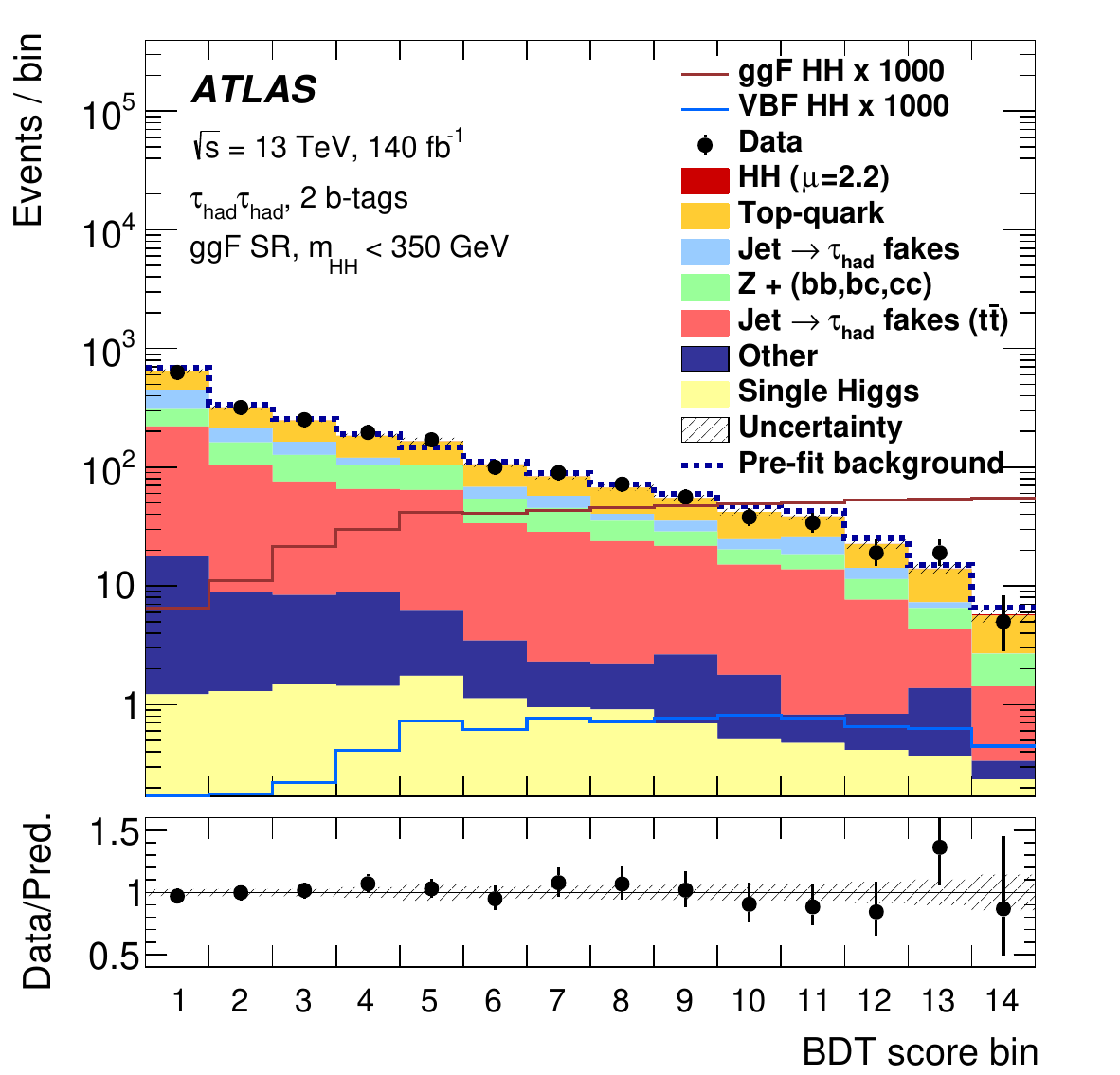}}
\subfigure[]{\includegraphics[width=0.32\linewidth]{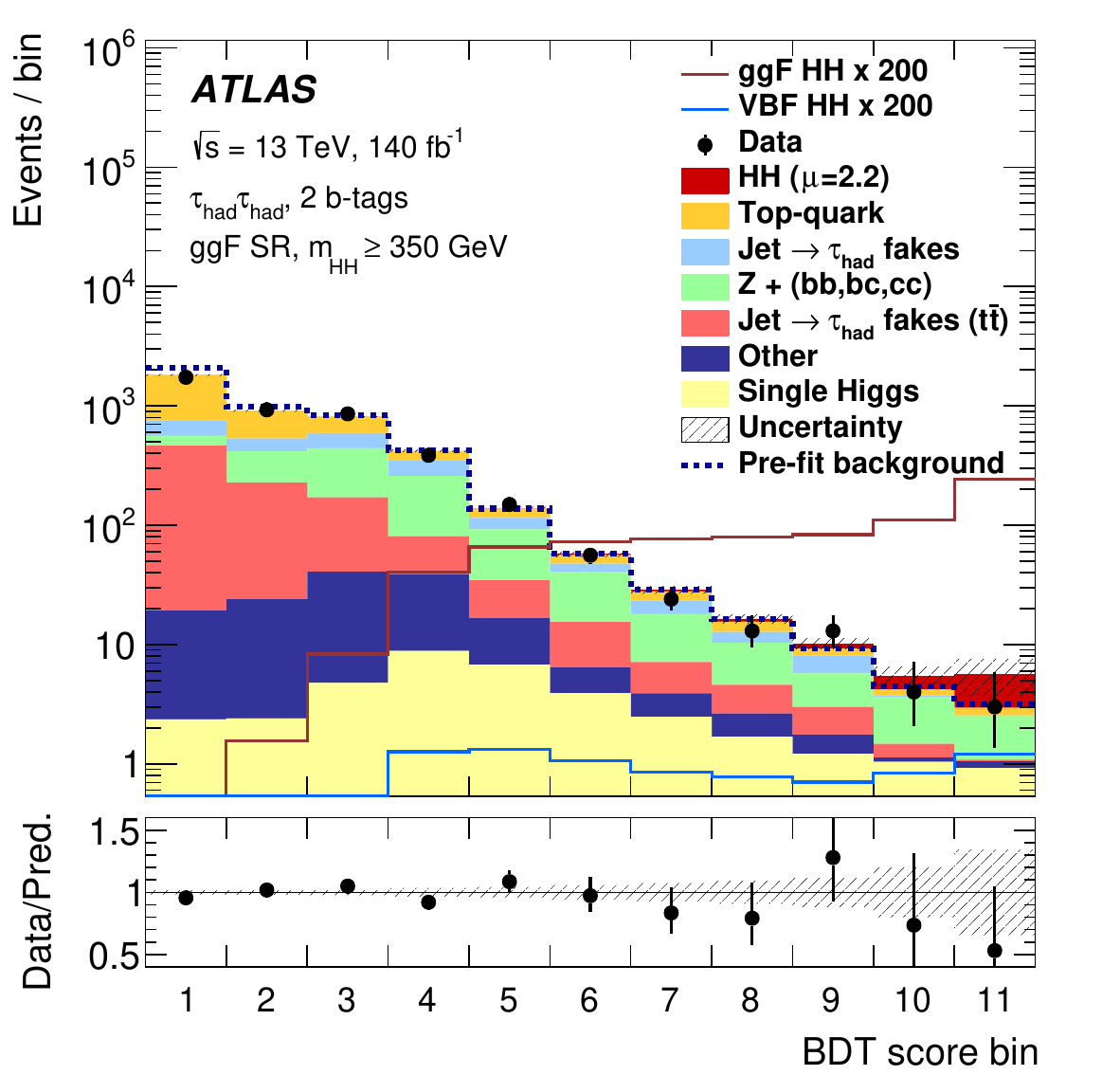}}
\subfigure[]{\includegraphics[width=0.32\linewidth]{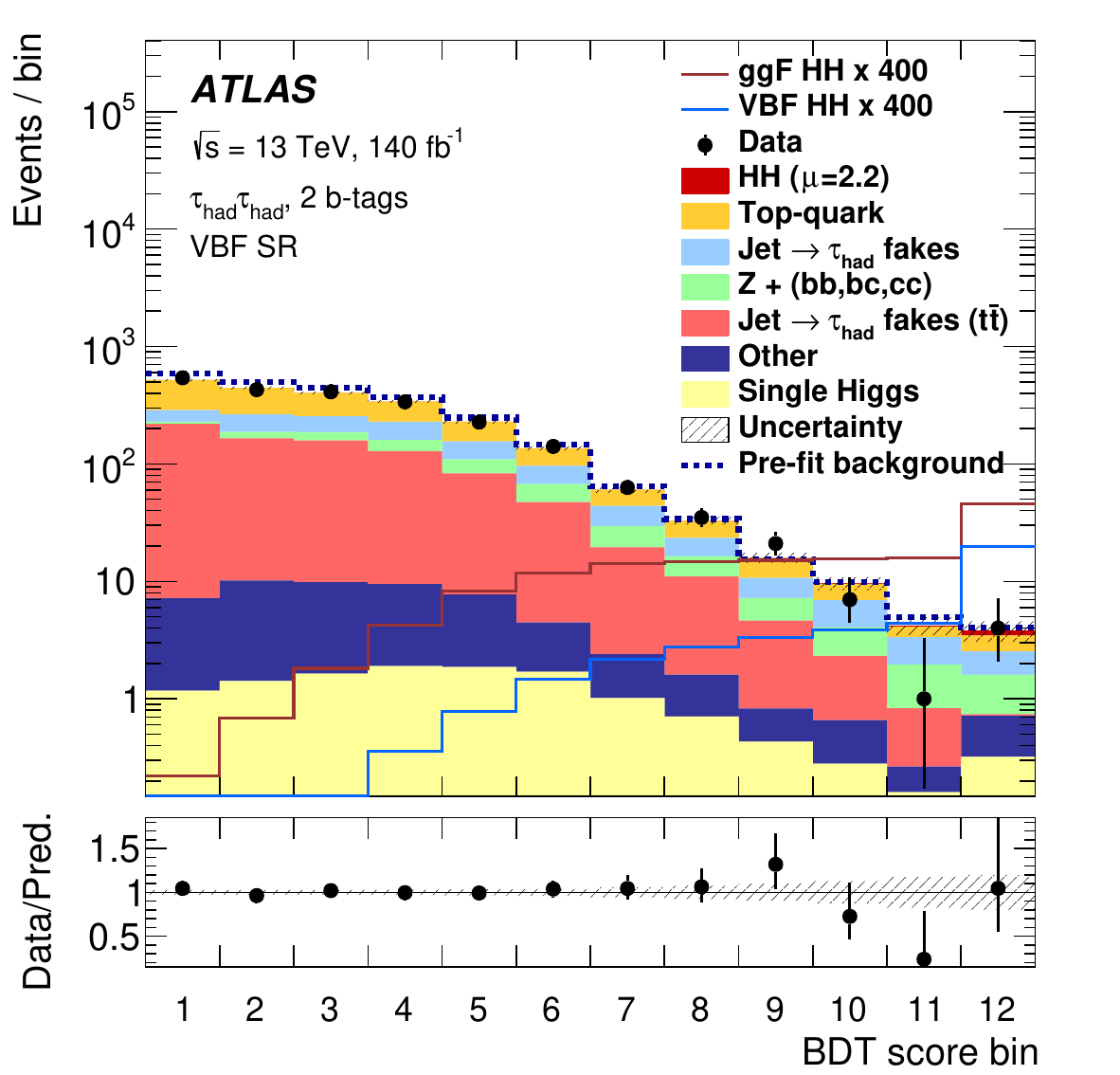}}
\subfigure[]{\includegraphics[width=0.32\linewidth]{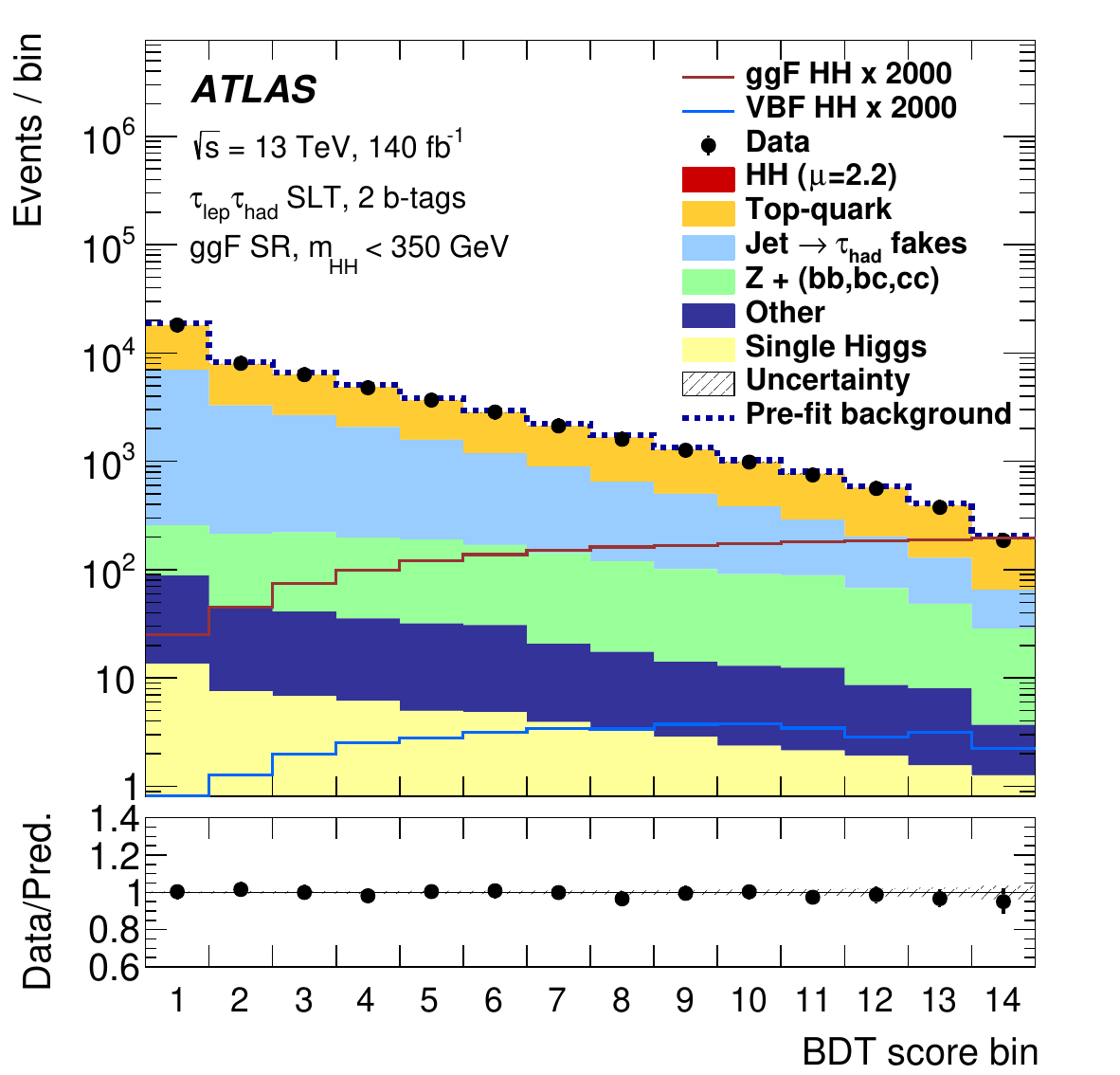}}
\subfigure[]{\includegraphics[width=0.32\linewidth]{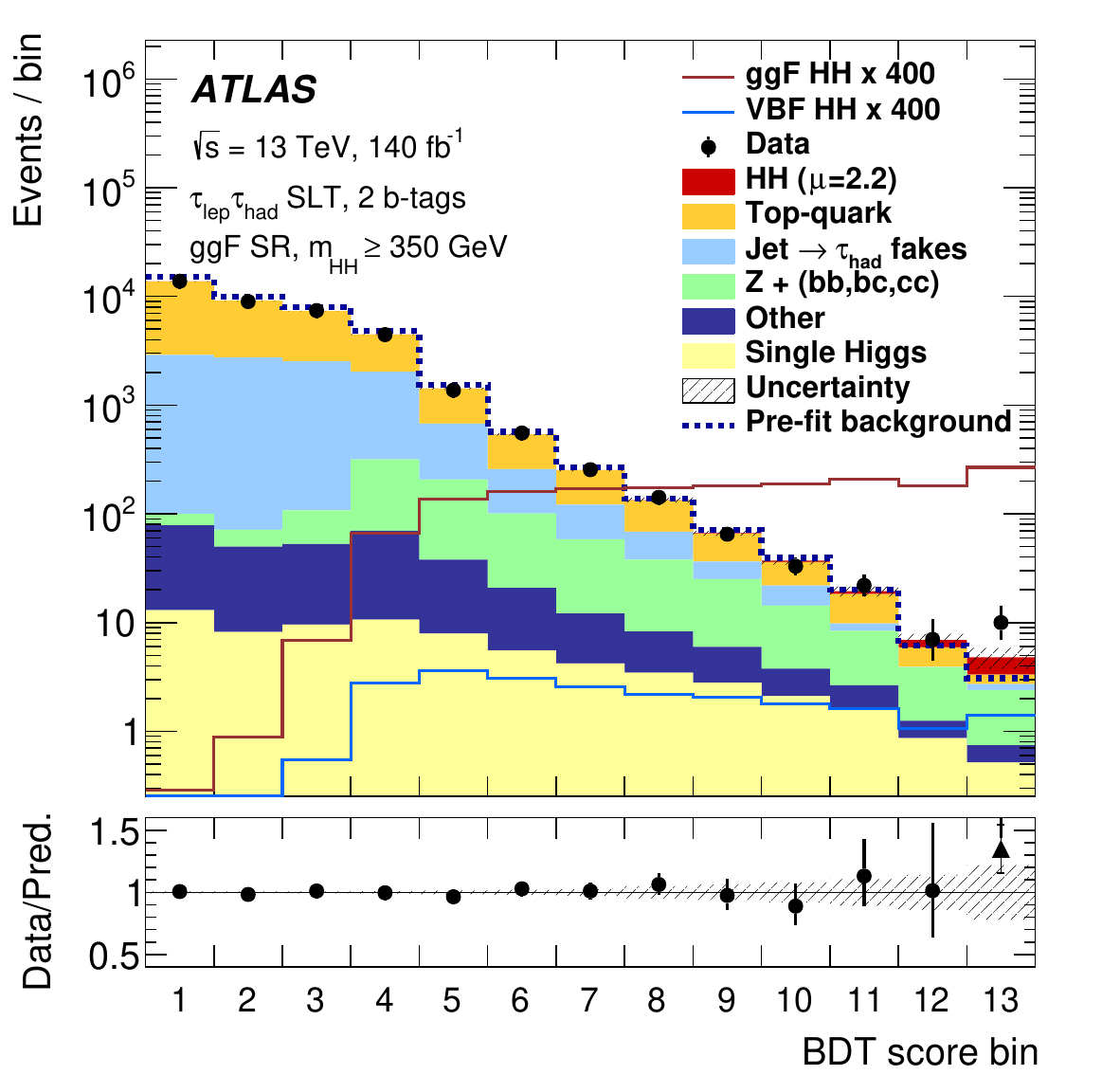}}
\subfigure[]{\includegraphics[width=0.32\linewidth]{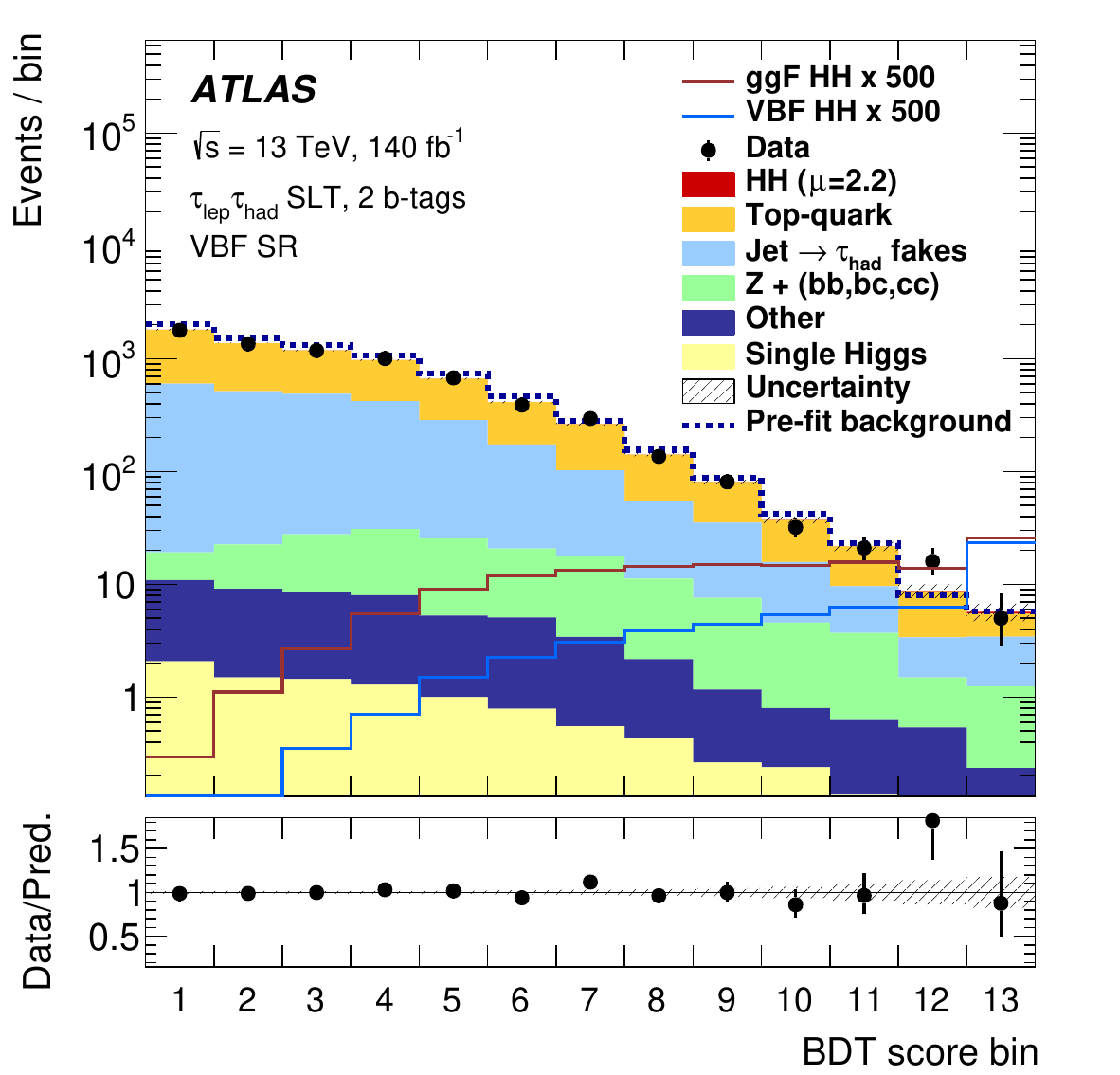}}\\
\subfigure[]{\includegraphics[width=0.32\linewidth]{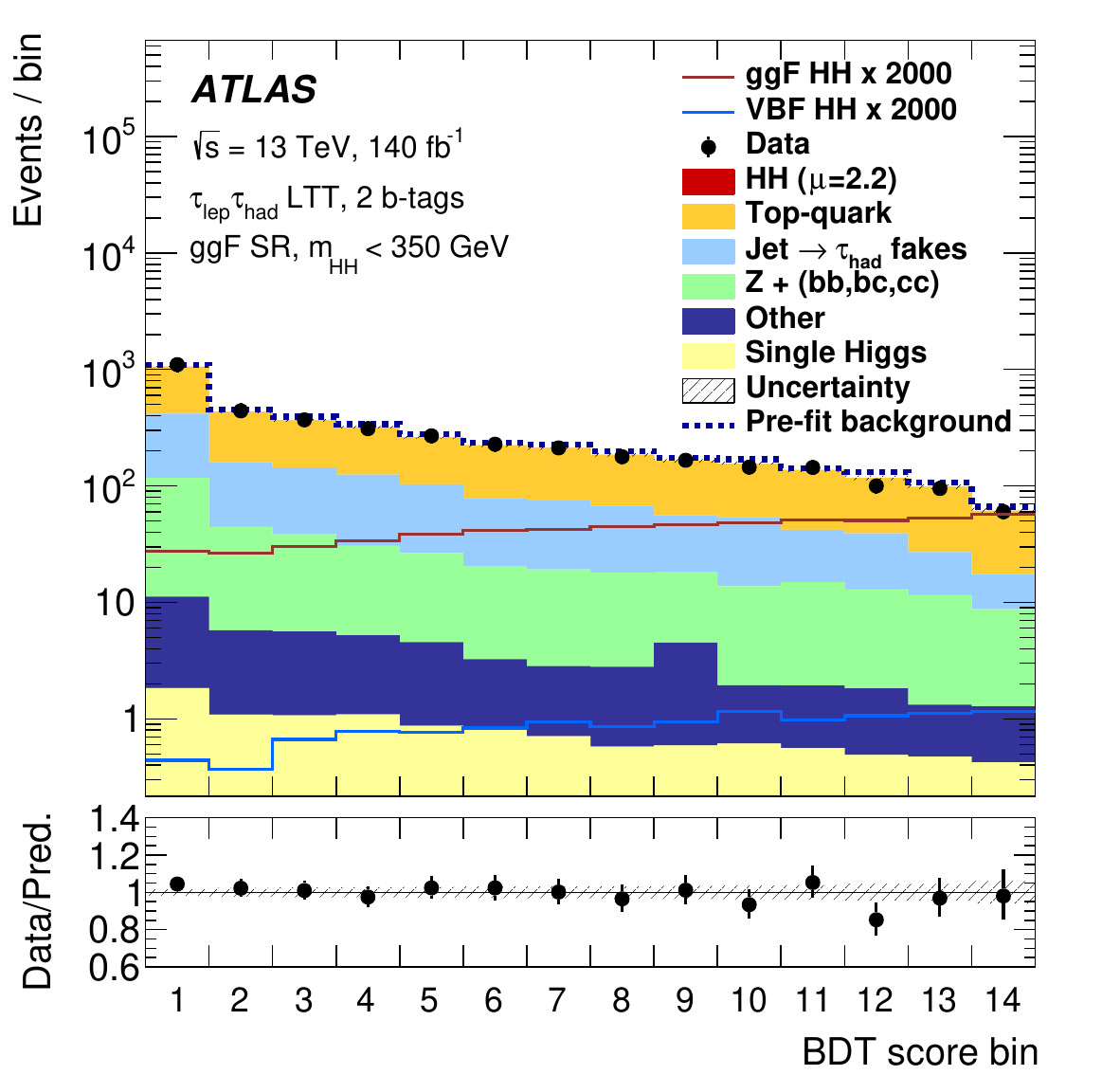}}
\subfigure[]{\includegraphics[width=0.32\linewidth]{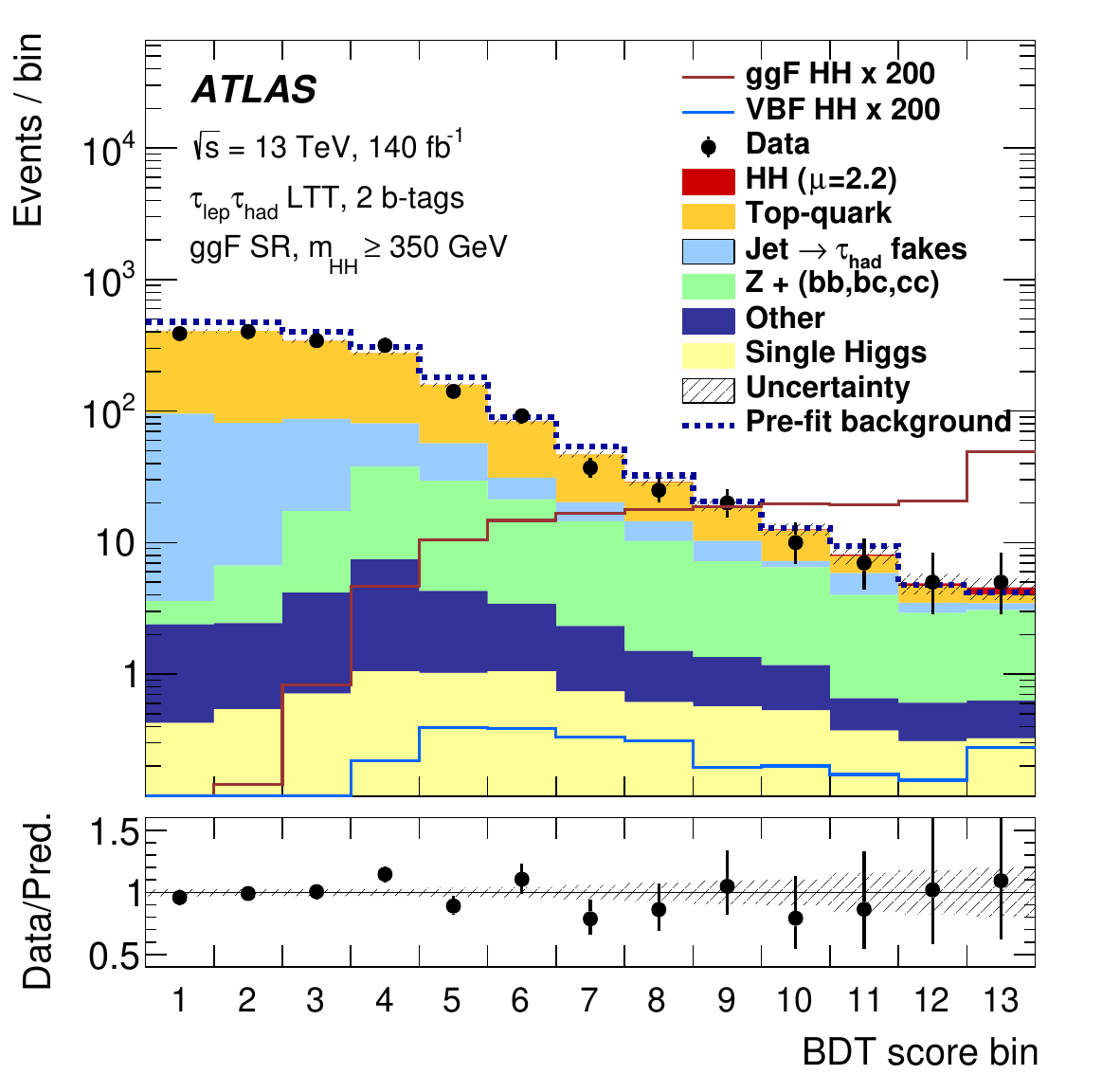}}
\subfigure[]{\includegraphics[width=0.32\linewidth]{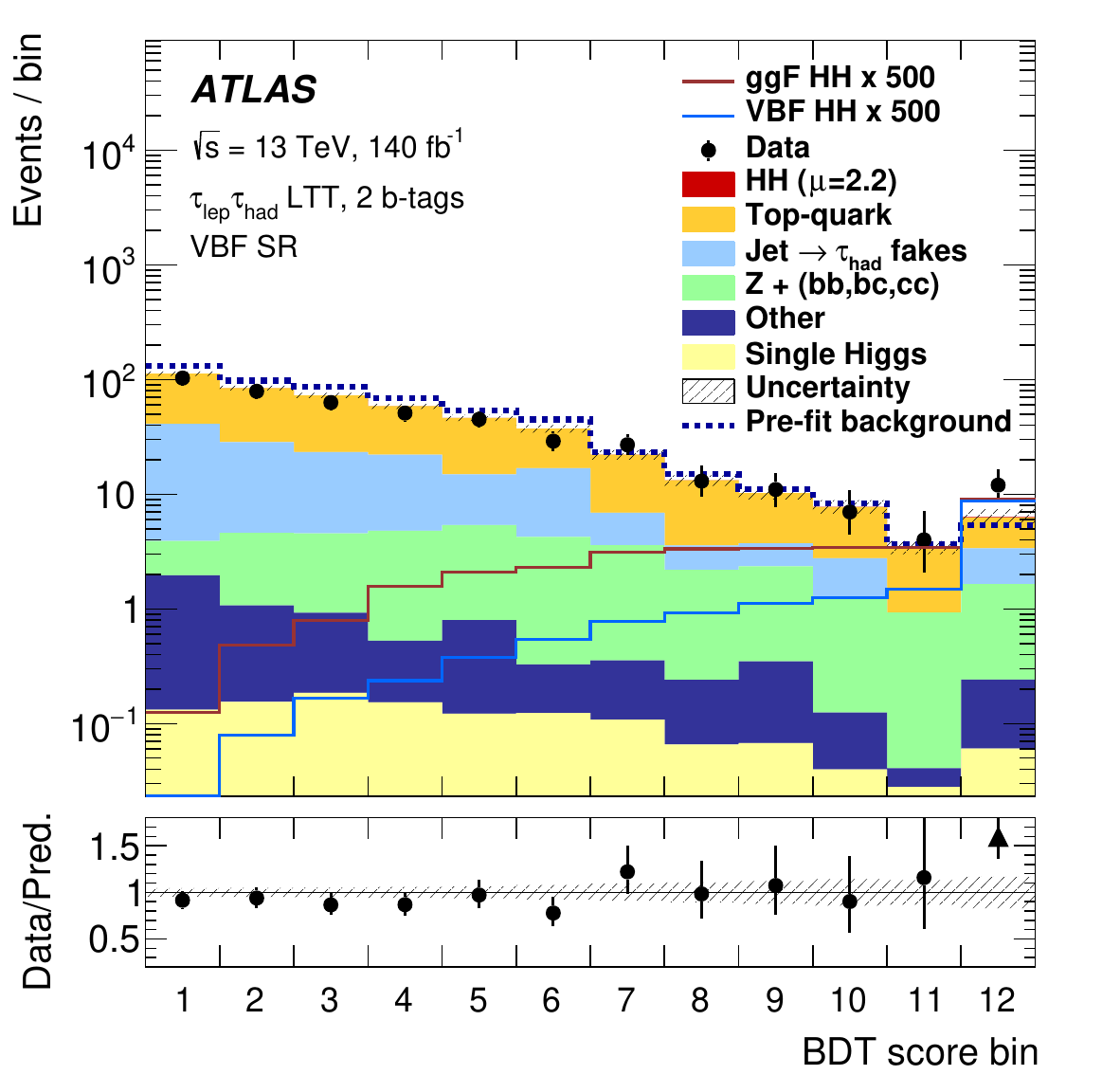}}\\
\caption{Predicted and observed distribution of the BDT score in~(left) the low-\mhh, (middle column) the high-\mhh and~(right) the VBF categories of~(top) the \hadhad, (middle row) the \lephad SLT and~(bottom) the \lephad LTT SRs.
The signal and background distributions are shown at post-fit level as obtained from the combined likelihood fit to data described in Section~\ref{sec:result}.
The background processes named `Other' contain contributions from $Z$-boson production in association with less than two jets initiated by $b$ or $c$ quarks, $W$-boson production, vector boson pair production, $t\bar{t}W$ and $t\bar{t}Z$ production processes.
The `HH' signal contribution is scaled to the fitted signal strength $\mu_{HH}$ from the combined likelihood fit times the SM expectation.
The ggF and VBF \HH signal distributions are overlaid and scaled to the factor indicated in the legend times the SM expectation.
The dashed histograms show the total pre-fit background.
The lower panels show the ratio of data to the total post-fit sum of signal and background, where the hatched bands show the statistical and systematic uncertainties of this prediction.
The BDT score distributions are shown with the binning used in the likelihood fit.
For visualisation purposes each bin is displayed with uniform width and an index labeling it.
}
\label{fig:bdtscores}
\end{figure}

The constraints on the coupling modifiers are determined using a profile likelihood ratio $\Lambda(\alpha, \theta)$ as the test statistic, computed from the likelihood function in the asymptotic approximation~\cite{Cowan:2010js}, where the POI $\alpha$ are the coupling modifiers $\kappa$.
The procedure adopted in Ref.~\cite{HDBS-2022-03} is used to set constraints on \kl and \kvv, expressed as 68\% and 95\% CIs.
Signal strength upper limits are derived using the CL$_\mathrm{s}$ method~\cite{Read:2002hq} with the procedure described in Ref.~\cite{HDBS-2018-40}.
Upper limits are set on the overall $\mu_{HH}$ and on the separate signal strength parameters $\mu_{\mathrm{ggF}} = \sigma_{\textup{ggF}} / \sigma_{\textup{ggF}}^{\textup{SM}} $ and $\mu_{\mathrm{VBF}} = \sigma_{\textup{VBF}} / \sigma_{\textup{VBF}}^{\textup{SM}} $.
The expected results are obtained with Asimov datasets~\cite{Cowan:2010js} generated from the likelihood function after setting all nuisance parameters to their maximum-likelihood estimate in the fit to the data and fixing the POIs to the values corresponding to the hypothesis under test.
The asymptotic results are found to agree within 7\% with the upper limits obtained using pseudo-experiments. %
Figure~\ref{fig:bdtscores} shows the BDT score distribution in the nine orthogonal categories, after performing the maximum-likelihood fit to data for the $L(\mu_{HH}, \theta)$ function.
Good agreement between the data and the prediction is found within the assessed uncertainties.

The maximum-likelihood estimator for the total \HH production signal strength is found to be $\hat{\mu}_{HH} = 2.2 \pm 1.7$ by the combined fit to data.
The uncertainty in the fitted signal strength is obtained from the variation of log-likelihood-based test statistic $\Lambda$ by one unit, and includes both statistical and systematic uncertainties.
The maximum-likelihood estimator for the unconstrained normalisation factor of the \ttbar and \ZHF backgrounds are measured at $0.96 \pm 0.03$ and $1.34 \pm 0.08$, respectively, by the combined fit to data.
An observed 95\% CL upper limit of 5.9 is set on $\mu_{\HH}$, to be compared with an expected limit of 3.3 in the background-only hypothesis ($\mu_{HH}=0$), corresponding to an observed (expected) significance with respect to the background-only hypothesis of $1.4~(0.75)~\sigma$.
From the simultaneous fit of $\mu_{\mathrm{ggF}}$ and $\mu_{\mathrm{VBF}}$, observed (expected) 95\% CL upper limits are $\mu_{\mathrm{ggF}}<5.8~(3.4)$ and $\mu_{\mathrm{VBF}}<91~(73)$, respectively, for each production mode, assuming that the signal strength parameters can vary independently for each production mode.
If $\mu_{\mathrm{VBF}}$ is fixed to the SM prediction, the observed (expected) 95\% CL upper limit is $\mu_{\mathrm{ggF}}<5.9~(3.4)$.
Similarly, if $\mu_{\mathrm{ggF}}$ is fixed to the SM prediction, the observed (expected) 95\% upper limit is $\mu_{\mathrm{VBF}}<93~(72)$.

\begin{figure}[!tbh]
\centering
\includegraphics[width=0.65\textwidth]{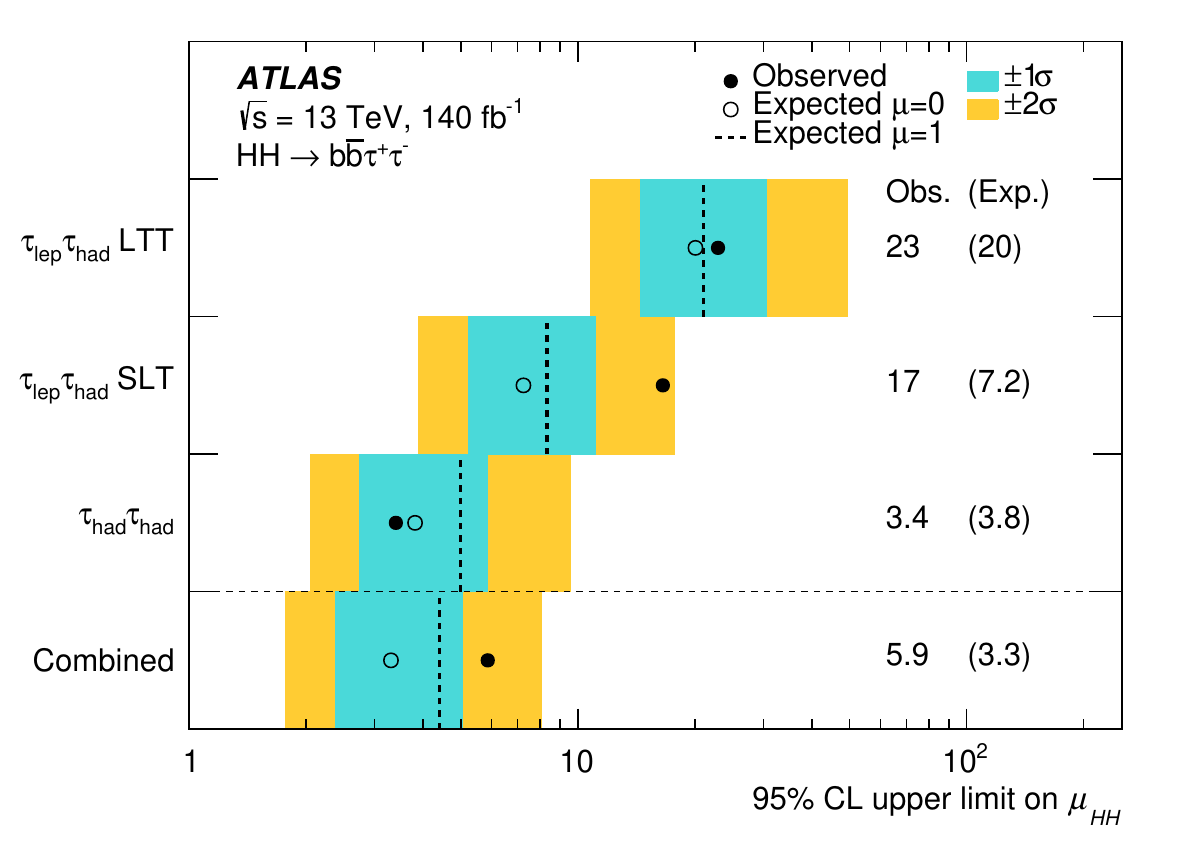}
\caption{Summary of observed (filled circles) and expected (open circles) 95\% CL upper limits on $\mu_{HH}$ from the fit of each individual channel and the combined fit in the background-only ($\mu_{HH}=0$) hypothesis.
The dashed lines indicate the expected 95\%~CL upper limits on $\mu_{HH}$ in the SM hypothesis ($\mu_{HH}=1$).
The inner and outer bands indicate the $\pm 1\sigma$ and $\pm 2\sigma$ variations, respectively, on the expected limit with respect to the background-only hypothesis due to statistical and systematic uncertainties.
}
\label{fig:limitSummary}
\end{figure}

Expected upper limits for the separate production mode signal strengths are derived with respect to the background-only hypothesis.
The signal strength upper limits are summarised in Table~\ref{tab:limits} and Figure~\ref{fig:limitSummary}, in each SR individually along with the combined fit.
The results for the individual SRs are obtained from the combined likelihood fit of the BDT score distributions in the categories of a single SR with the \mll distribution from the dedicated CR\@.
The observed limit on $\mu_{HH}$ from the combined fit is looser than the expected one as a result of an excess in the \lephad SLT SR, in the high-\mhh category.
This excess corresponds to a local significance of 2.3~$\sigma$ with respect to the SM hypothesis ($\mu_{HH}=1$), and a local significance of 2.7~$\sigma$ with respect to the background-only hypothesis, as obtained from the individual fit of the \lephad SLT SR.
\FloatBarrier

\begin{table}[!tbh]
\setlength
\extrarowheight{2.55pt}
\centering
\caption{Observed and expected 95\% CL upper limits on $\mu_{HH}$, $\mu_{\mathrm{ggF}}$ and $\mu_{\mathrm{VBF}}$ from the individual SR likelihood fits as well as the combined results.
The $\mu_{\mathrm{ggF}}$ and $\mu_{\mathrm{VBF}}$ limits are quoted both from the results of the simultaneous fit of both signal strengths (central column), and from independent fits for the individual production modes, assuming the other to be as predicted by the SM.
The uncertainties quoted on the combined expected upper limits correspond to the $1\sigma$ uncertainty band. \protect\label{tab:limits}}
\begin{tabular}{ccc:cc:cc}
\toprule
& & $\mu_{HH}$ & $\mu_{\mathrm{ggF}}$ & $\mu_{\mathrm{VBF}}$ & $\mu_{\mathrm{ggF}}$ ($\mu_{\mathrm{VBF}}$=1) & $\mu_{\mathrm{VBF}}$ ($\mu_{\mathrm{ggF}}$=1)\\
\midrule
\multirow{2}{*}{\hadhad} & observed & 3.4 & 3.6 & 87 & 3.5 & 80 \\[2.5pt]
& expected & 3.8 & 3.9 & 102  & 3.9 & 99 \\[2.5pt]
\cdashline{2-7}
\multirow{2}{*}{\lephad SLT} & observed & 17 & 17 & 136 & 17 & 158 \\[2.5pt]
& expected & 7.2 & 7.4 & 129  & 7.4 & 127 \\[2.5pt]
\cdashline{2-7}
\multirow{2}{*}{\lephad LTT} & observed & 23 & 18 & 765 & 22 & 733 \\[2.5pt]
& expected & 20 & 21 & 359 & 20 & 350 \\[2.5pt]
\cline{2-7}
\multirow{2}{*}{Combined} & observed & 5.9 & 5.8 & 91 & 5.9 & 93 \\[2.5pt]
& expected & $3.3^{+1.7}_{-0.9}$ & $3.4^{+1.8}_{-1.0}$ & $73^{+32}_{-21}$ & $3.4^{+1.8}_{-0.9}$ & $72^{+32}_{-20}$ \\[2.5pt]
\bottomrule
\end{tabular}
\end{table}

\begin{figure}[!tbh]
\centering
\subfigure[]{\includegraphics[width=0.45\textwidth]{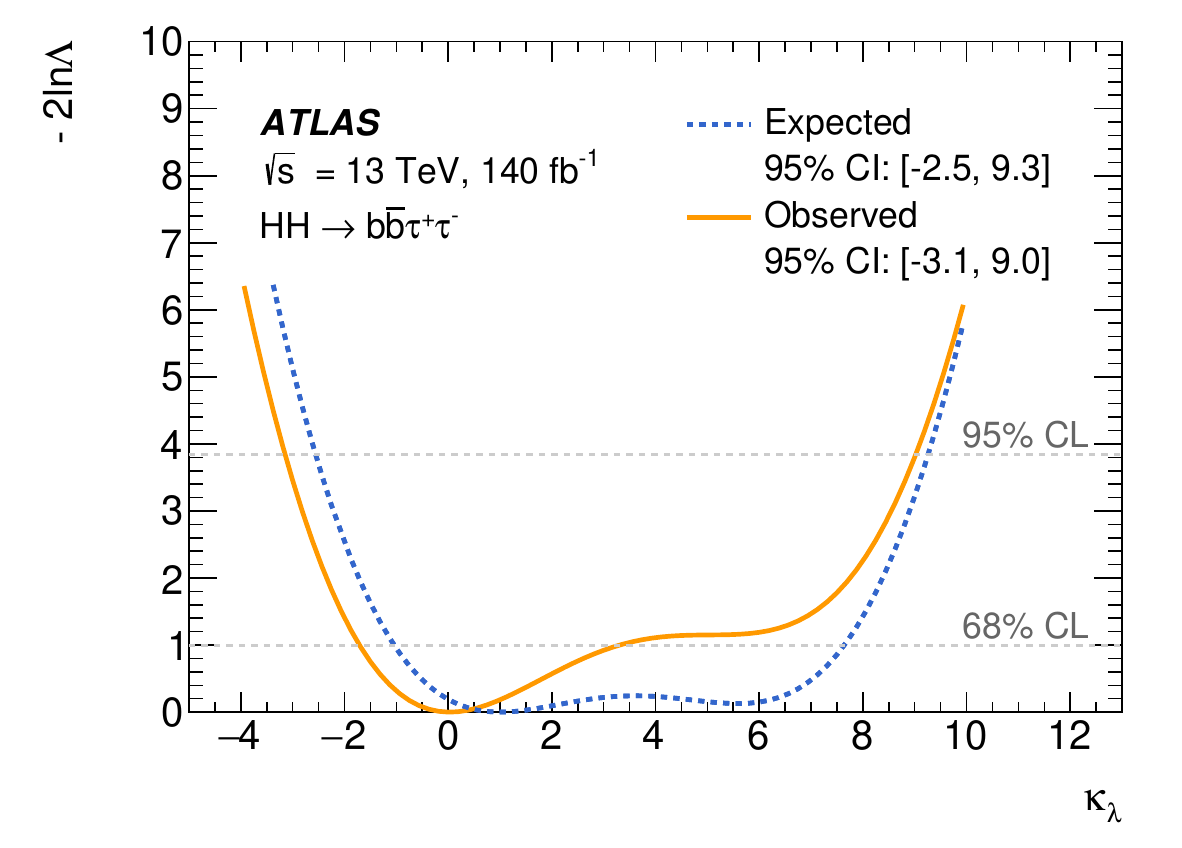}\label{sfg:klambdaObsExp}}
~
\subfigure[]{\includegraphics[width=0.45\textwidth]{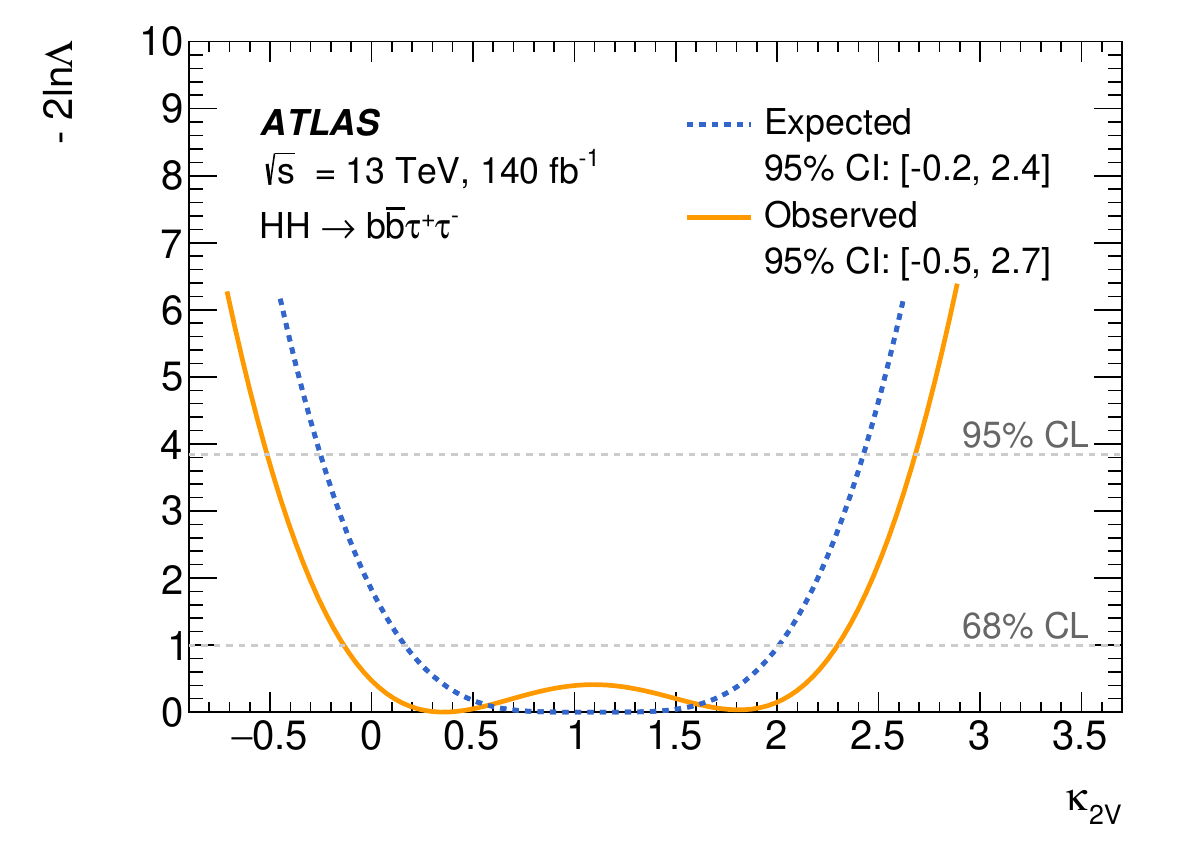}}
\caption{Values of $-2\ln\Lambda$ for different~(a) \kl and~(b) \kvv hypotheses obtained from fits to the data (solid orange) and an Asimov dataset (dashed blue) constructed under the SM hypothesis.
In each case, all coupling modifiers but the scanned parameter are fixed to their SM values.
}
\label{fig:kappa1d}
\end{figure}

\begin{figure}[!tbh]
\centering
\includegraphics[width=0.65\textwidth]{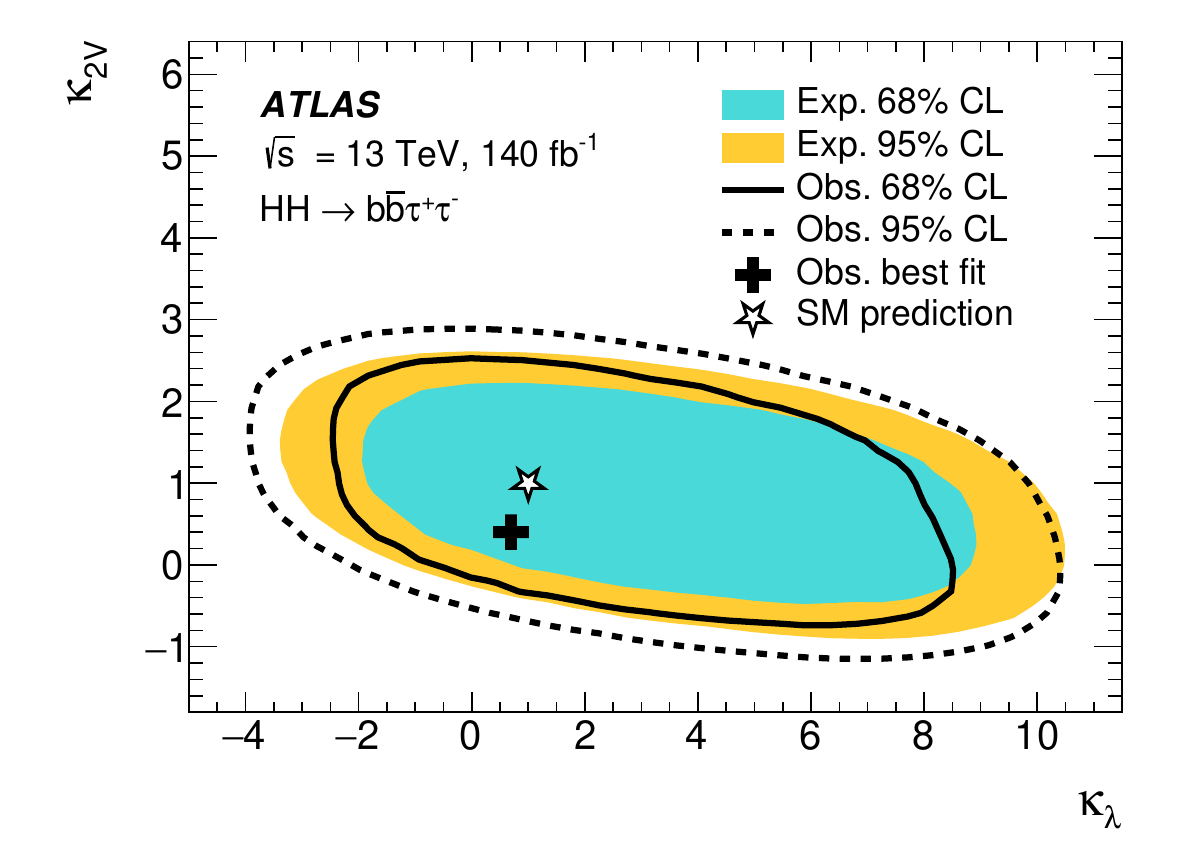}
\caption{Likelihood contours at 68\% (solid line) and 95\% (dashed line) CL in the (\kl, \kvv) parameter space, when all other coupling modifiers are fixed to one.
The corresponding expected contours are shown by the inner and outer shaded regions.
The SM prediction is indicated by the star, while the best-fit value is denoted by the black cross.}
\label{fig:kappa2d}
\end{figure}

The observed and expected values of $-2\ln\Lambda$ as a function of the coupling modifiers \kl and \kvv are shown in Figure~\ref{fig:kappa1d}, under the hypothesis that all other coupling modifiers are equal to their SM predictions. %
The combined fit allows to set observed (expected) 95\% CIs of $\kl\in [-3.1, 9.0]~([-2.5, 9.3])$ (assuming $\kvv=1$) and $\kvv\in [-0.5, 2.7]~([-0.2, 2.4])$ (assuming $\kl=1$).
Additional constraints are set on \kl and \kvv assuming that both coupling modifiers can vary simultaneously.
The resulting observed and expected two-dimensional 68\% and 95\% contours are shown in Figure~\ref{fig:kappa2d}.
The observed and expected constraints on \kl reported in this paper are affected by the issue concerning the ggF \HH prediction for BSM scenarios in \POWHEG reported in the erratum of Ref.~\cite{Heinrich:2022idm} and do not include the changes to resolve this issue described in Ref.~\cite{fixedPow:2023}.
If the ggF \HH signal yields in the analysis categories are scaled based on the ratio of the predicted differential ggF \HH cross-sections with and without the changes described in Ref.~\cite{fixedPow:2023}, the width of the 95\% CI on \kl changes by less than 5\%.

As in Ref.~\cite{HDBS-2018-40}, the analysis sensitivity is primarily limited by the statistical uncertainty of the data.
The leading systematic uncertainty in the measurement of $\mu_{HH}$ is the uncertainty in the ggF \HH production cross-section arising from variations of the QCD scales and the top-quark mass scheme.
The next leading sources of uncertainty are the statistical precision of the background MC samples and the uncertainty related to the interference between the $Wt$ and $t\bar{t}$ processes.
The combined impact of all sources of systematic uncertainties leads to an increase in the expected upper limits on the signal strength $\mu_{HH}$ by 23\% and to a widening of the expected 95\% CI for \kl and \kvv by 9\% and 2\%, respectively, with respect to the case in which systematic uncertainties are neglected (excluding the \ttbar and \ZHF floating normalisation and MC statistical uncertainties).

Based on a consistent statistical procedure for evaluating the 95\% CIs and upper limits as described at the beginning of this section, these results can be compared with the previous analysis of Ref.~\cite{HDBS-2018-40}.
The approach presented in this paper leads to an increase (reduction) of the observed (expected) upper limit on the signal strength $\mu_{HH}$ by 25\% (15\%), along with new results for the upper limits on the separate ggF and VBF \HH production mode signal strengths.
The width of the observed (expected) CI for \kl is reduced by $<1\%$ ($11$\%) and the width of the observed (expected) CI for \kvv is reduced by 2\% (19\%) compared with the previous analysis.

The compatibility, considering statistical and systematic uncertainties, between the upper limit at 95\% CL on the signal strength $\mu_{HH}$ from this study and that of Ref.~\cite{HDBS-2018-40} is evaluated using a bootstrap technique~\cite{PhysRevD.39.274} separately for the independent SR fits and for the combined fit.
The compatibility between the two results is at the level of 1$\sigma$ for the individual fit of the \hadhad SR, of 2.5$\sigma$ for the individual fit of the \lephad SLT SR, and of 0.1$\sigma$ for the individual fit of the \lephad LTT SR\@.
The compatibility is at the level of 0.8$\sigma$ for the combined fit.

%
%

%
%

%
%
%
%
%
%
%
%
%
%


\section{Effective field theory interpretation}

\label{sec:eft}
Effective field theory approaches can be employed to parameterise effects of new particles and anomalous couplings on the \HH production rates and Higgs boson decay rates, assuming that the energy scale of the underlying BSM processes is far too high to be probed directly.
Interpretations of the results in the context of the Higgs Effective Field Theory (HEFT) and SM Effective Field Theory (SMEFT) are presented in this section, following closely the procedure in Refs.~\cite{HDBS-2019-29, HDBS-2021-10-journal}.
The SMEFT interpretation is based on the effective Lagrangian
\begin{align*}
\mathcal{L}_{\mathrm{SMEFT}} = \mathcal{L}_{\mathrm{SM}} + \frac{1}{\Lambda_{\mathrm{EFT}}^2} \sum\limits_k c_k^{(6)} O_k^{(6)}\,,
\end{align*}
where $\mathcal{L}_{\mathrm{SM}}$ is the SM Lagrangian and $\Lambda_{\mathrm{EFT}}$ is the energy scale of the new physics phenomena, which is set to \qty{1}{\TeV}.
The $c_k^{(6)}$ are the Wilson coefficients corresponding to an operator $O_k^{(6)}$ of mass dimension 6.
The Warsaw basis~\cite{Grzadkowski:2010es} is chosen for the dimension-6 operators, and constraints are placed on the Wilson coefficients \cH and \cHbox.
They correspond to the operators $(\phi^\dagger \phi)^3$ and $(\phi^\dagger \phi)\Box(\phi^\dagger \phi)$, respectively.
While the VBF \HH contribution is affected by variations of \cH and \cHbox, the SMEFT preserves the Higgs doublet structure of the SM, which implies cancellations between the LO diagrams \ref{sfg:vvhh} and \ref{sfg:vvh}.
Therefore, the change in the expected VBF \HH contribution is small, and the process is ignored for the SMEFT interpretations, as it is dominated by the ggF \HH contribution in all considered coupling scenarios.

For the HEFT interpretation, two studies are conducted. %
First, cross-section limits are placed on the seven \mhh shape benchmark scenarios defined in Ref.~\cite{alasfar:2023}.
As described in Ref.~\cite{capozi:2020}, these benchmark scenarios have been determined using a clustering algorithm that identifies distinct shapes of the \mhh spectrum that can be obtained by varying five Wilson coefficients in the HEFT framework.
In addition to the study of these benchmark scenarios, the Wilson coefficients \ctthh and \cgghh are considered, which quantify the strengths of the effective $t\bar{t}HH$ and $ggHH$ couplings, respectively.
The Wilson coefficient of the only considered anomalous coupling that affects VBF \HH production at NLO precision is \chhh, which is equivalent to \kl.
Due to the much smaller overall expected yield compared with ggF \HH production, and the decorrelation of the two production modes through the event categorisation, the impact of VBF \HH contribution on the \mhh shape benchmark limits is expected to be negligible, even when taking \chhh variations into account.
Hence, it is ignored in the following, i.e., $\mu_{\mathrm{VBF}}$ is set to 0.

Signal predictions for ggF \HH production assuming different effective coupling strengths are obtained from the SM sample.
For this purpose, a reweighting method is used that assigns each event a weight corresponding to the ratio of the differential NLO cross-section (in \mhh) predicted by the SM and the hypothesis of interest.
These weights are taken from Ref.~\cite{alasfar:2023}. %
Possible effects of the anomalous couplings on the Higgs boson branching fractions are neglected.
The SMEFT coefficients can also affect single Higgs boson processes.
To take these effects on production cross-sections and branching fractions into account, the procedure in Ref.~\cite{ATLAS:2024lyh} is adopted.
For a continuous parameterisation of the effect of the Wilson coefficients on the predicted ggF \HH distributions, the linear combination method described in Ref.~\cite{ATL-PHYS-PUB-2019-007} is applied to the ggF \HH cross-section dependence on the respective Wilson coefficients.
The basis samples for the Wilson coefficients are chosen to be close to the 95\% CL limits in order to reduce the effect of large cancellations between the basis samples on the obtained results.
The values of the couplings $(\cgghh, \ctthh, \chhh)$ that define the ten samples making up the basis for measuring the HEFT Wilson coefficients are $(0, 0, 1)$, $(-0.3, 0, 1)$, $(0.4, 0, 1)$, $(0, -0.2, 1)$, $(0, 0.7, 1)$, $(0, 0, -2.5)$, $(0, 0, 9)$, $(0.4, 0, -2.5)$, $(0, 1, -10)$, $(0.4, -0.2, 1)$.
The six values of $(\cH, \cHbox)$ used for the SMEFT interpretation are $(0, 0)$, $(-17, 0)$, $(7, 0)$, $(0, -7)$, $(0, 13)$, $(7, -7)$.

Uncertainties on the ggF \HH event yield from the reweighting procedure are derived using comparisons to alternative MC samples generated with different assumed coupling values.
They are produced for the seven HEFT \mhh shape benchmarks and for $c_{gghh/tthh}\in \{-0.5, 0.5, 1\}$ using \POWHEGBOX[v2] interfaced with \Pythia[8.244] for parton showering and hadronisation. %
The uncertainty assessment for variations in \cH is based on the LO relation $\kappa_\lambda = 1 - 2\frac{v^4}{m_H^2\Lambda_{\mathrm{EFT}}^2}c_H + 3\frac{v^2}{\Lambda_{\mathrm{EFT}}^2}\left(\cHbox - \frac{1}{4} c_{HD}\right)$~\cite{alasfar:2023}, which allows for a comparison to a $\kl = 10$ sample. Here, $c_{HD}$ is the Wilson coefficient corresponding to the dimension-6 operator $(\phi^\dagger D_\mu \phi)^*(\phi^\dagger D^\mu \phi)$. %
For \cHbox, alternative samples with $\cHbox \in \{-9.5, 7\}$ are used. %
Additionally, the signal modelling uncertainties are evaluated for each signal configuration, as described in Section~\ref{sec:systs}.
For scans of Wilson coefficients, the modelling uncertainties are computed for all signal configurations used in the linear combination approach.
The maximal obtained deviation is taken as the final value for each source of uncertainty.

\begin{figure}[!tbh]
\centering
\includegraphics[width=.65\textwidth]{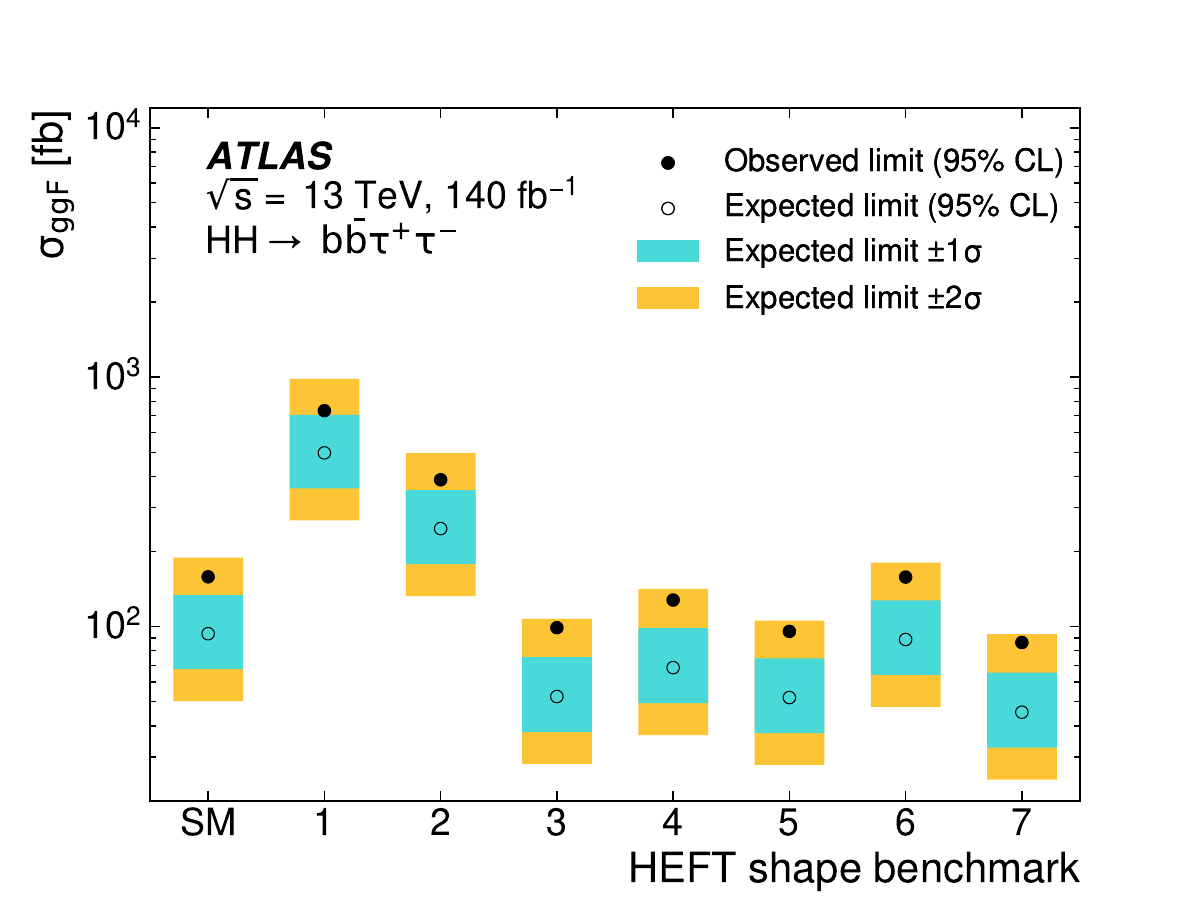}
\caption{Observed (filled circles) and expected (open circles) 95\% CL upper limits on the ggF \HH production cross-section with respect to the background-only hypothesis in the SM and each of the seven HEFT shape benchmarks obtained from the combined fit.
The inner and outer bands show the $\pm1\sigma$ and $\pm2\sigma$ variations on the expected upper limit.
The contribution from VBF \HH production is neglected for this result.
}
\label{fig:bmLimits}
\end{figure}

The 95\% CL upper limits on the ggF \HH cross-section assuming the different benchmark \mhh shapes are summarised in Figure~\ref{fig:bmLimits}.
The following changes are made in the fit model with respect to the measurement of signal strength limits described in Section~\ref{sec:result}.
The ggF \HH prediction is replaced depending on the benchmark scenario, while the contribution from VBF \HH production is neglected.
Moreover, as this is a cross-section measurement, no uncertainties in the ggF \HH production cross-section are taken into account.
Overall, it is observed that lower average signal \mhh values, as represented by benchmarks 1 and 2, lead to weaker constraints.

The Wilson coefficients are probed with the same profile likelihood ratio scan as for \kl and \kvv in Section~\ref{sec:result}.
The one-dimensional constraints on the HEFT and SMEFT Wilson coefficients are summarised in Table~\ref{tab:eft1d}.

The observed and expected constraints on the ggF \HH cross-section limits assuming the HEFT shape benchmarks shown in Figure~\ref{fig:bmLimits} and the CIs on individual Wilson coefficients reported in Table~\ref{tab:eft1d} are affected by the issue concerning the ggF \HH prediction for BSM scenarios in \POWHEG reported in the erratum of Ref.~\cite{Heinrich:2022idm} and do not include the changes to resolve this issue described in Ref.~\cite{fixedPow:2023}.
If the ggF \HH signal yields in the analysis categories are scaled based on the ratio of the predicted differential ggF \HH cross-sections with and without the changes described in Ref.~\cite{fixedPow:2023}, the expected limits and CIs changes by less than 10\%.
\begin{table}[!tbh]
\centering
\caption{Observed and expected 95\% CIs on HEFT and SMEFT Wilson coefficients, obtained from one-dimensional scans of $-2\ln\Lambda$ in fits to the observed data and an Asimov dataset constructed under the SM hypothesis.
In each case, only the scanned Wilson coefficient is varied while all others are fixed to their respective SM predicted value.
The contribution from VBF \HH production is neglected for this result.
The SM value for all considered Wilson coefficients is 0, hence no sign of BSM physics is found.
}
\label{tab:eft1d}
\begin{tabular}{ccc}
\toprule
Wilson coefficient & Observed 95\% CI & Expected 95\% CI\\
\midrule
\cgghh&[\num{-0.51}, \num{0.58}]&[\num{-0.42}, \num{0.44}]\\
\ctthh&[\num{-0.40}, \num{0.84}]&[\num{-0.32}, \num{0.72}]\\
\midrule
\cH&[\num{-19.4}, \num{10.0}]&[\num{-19.1}, \num{8.6}]\\
\cHbox&[\num{-12.6}, \num{11.6}]&[\num{-8.5}, \num{11.1}]\\
\bottomrule
\end{tabular}
\end{table}

\begin{figure}[!tbh]
\centering
\subfigure[]{\includegraphics[width=.47\textwidth]{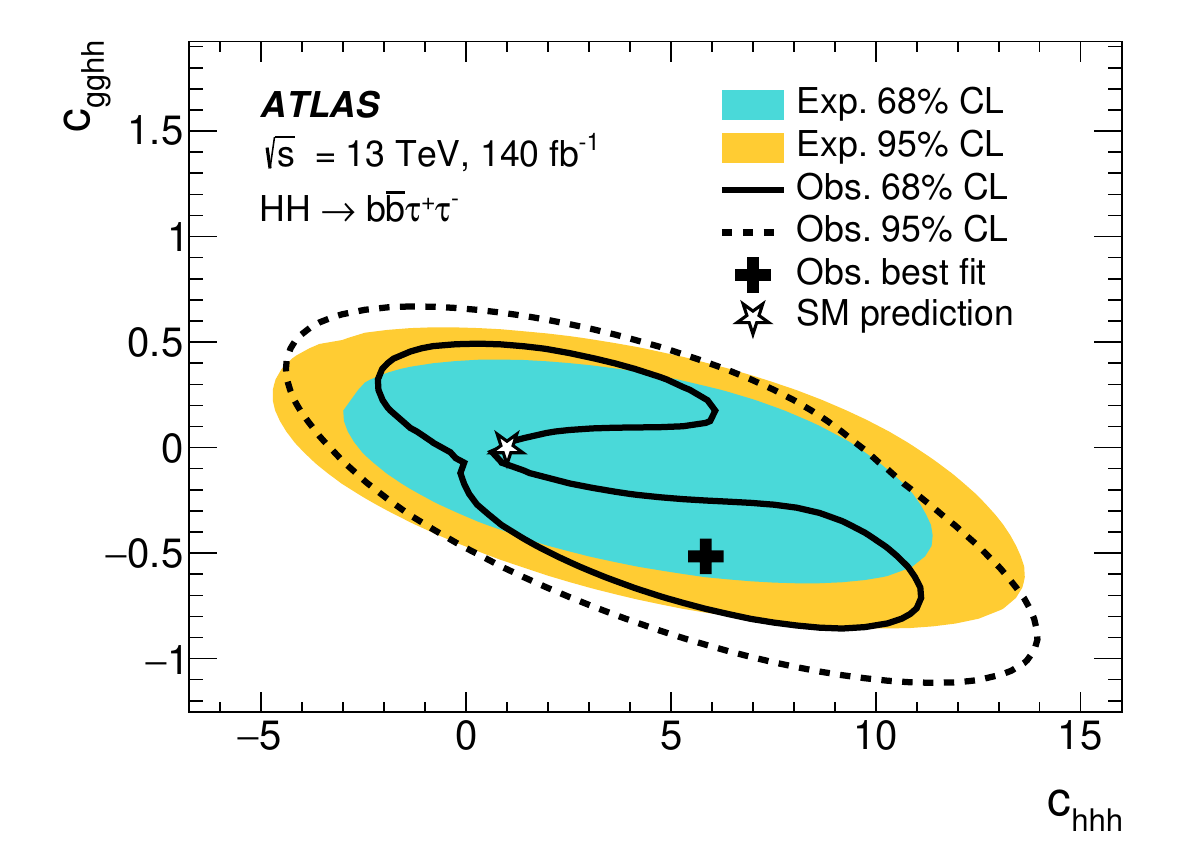}\label{sfg:gghh2d}}
\subfigure[]{\includegraphics[width=.47\textwidth]{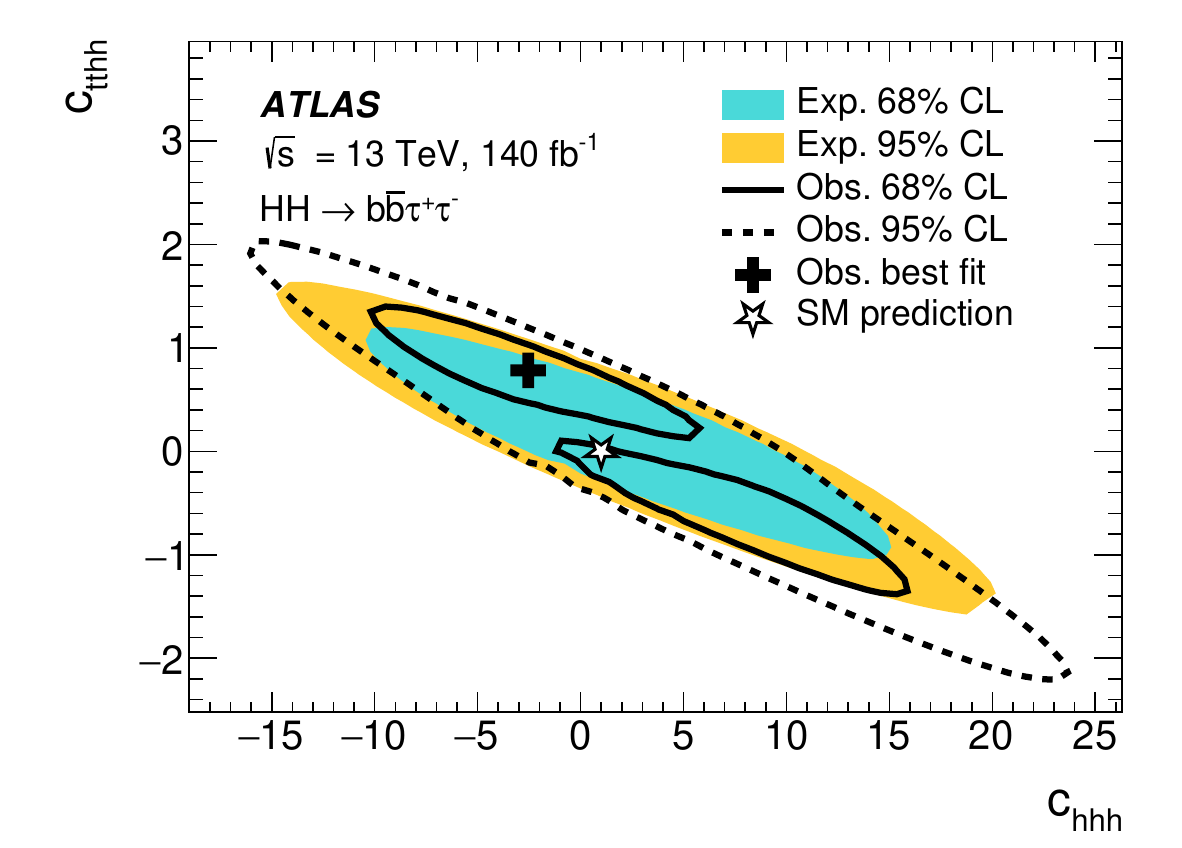}\label{sfg:tthh2d}}
\caption{Likelihood contours at 68\% (solid line) and 95\% (dashed line) CL in~(a) the (\chhh, \cgghh) and~(b)the (\chhh, \ctthh) parameter space, when all other Wilson coefficients are fixed to their SM values.
The corresponding expected contours are shown by the inner and outer shaded regions.
The SM prediction is indicated by the star, while the best-fit value is denoted by the black cross.}
\label{fig:heft2d}
\end{figure}

\begin{figure}[!tbh]
\centering
\includegraphics[width=.47\textwidth]{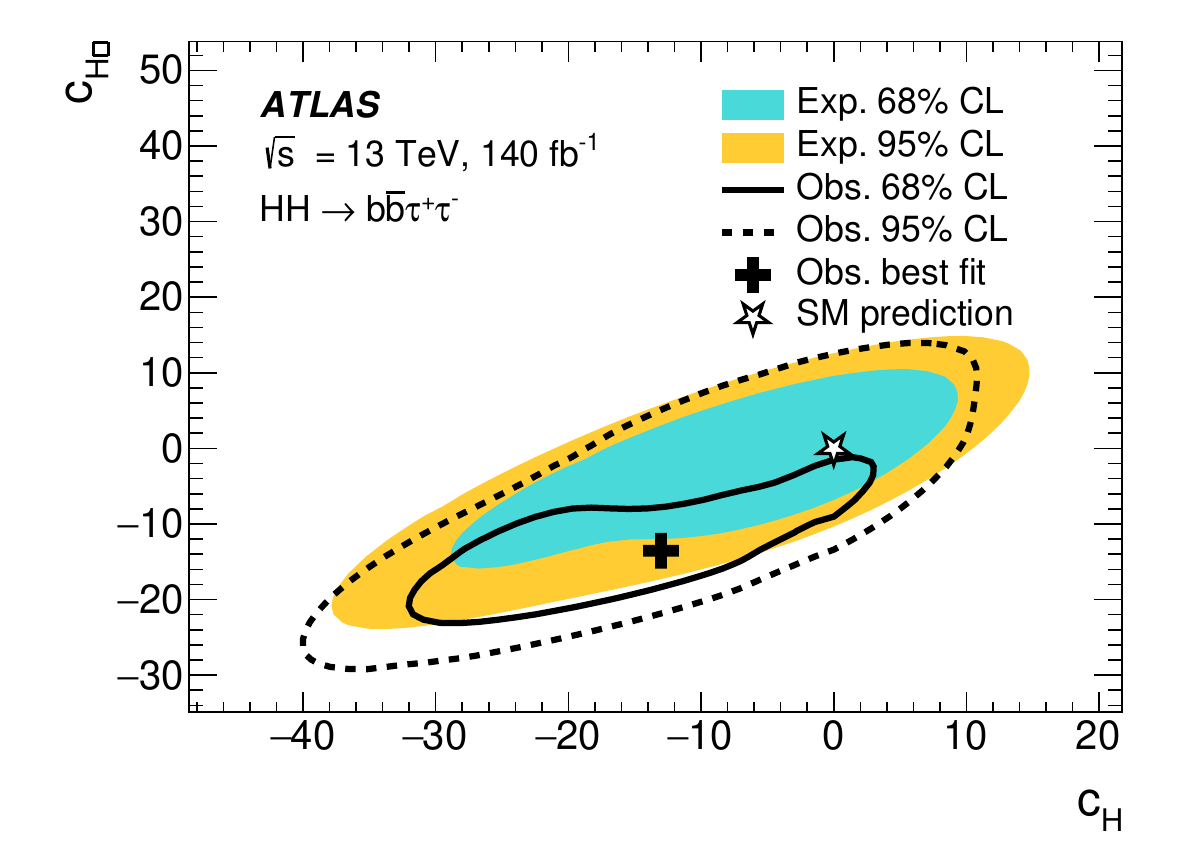}
\caption{Likelihood contours at 68\% (solid line) and 95\% (dashed line) CL in the (\cH, \cHbox) parameter space, when all other Wilson coefficients are fixed to their SM values.
The corresponding expected contours are shown by the inner and outer shaded regions.
The SM prediction is indicated by the star, while the best-fit value is denoted by the black cross.}
\label{fig:smeft2d}
\end{figure}

Two-dimensional 68\% and 95\%~CL likelihood contours are measured for the HEFT Wilson coefficients \cgghh and \ctthh, respectively, as a function of \chhh.
They are shown in Figure~\ref{fig:heft2d}.
In addition, two-dimensional contours are measured for the SMEFT coefficients \cH and \cHbox, as shown in Figure~\ref{fig:smeft2d}.
The observed non-parabolic features in the two-dimensional contours, such as the double-minimum structure in Figure~\ref{sfg:tthh2d}, stem from different best-fit Wilson coefficient values in the \lephad~SLT and \hadhad high-\mhh categories, which drive the sensitivity of the combined fit.
The excess in the SLT high-\mhh category leads to a 68\%~CL rejection of the region between the two 68\%~CL contours, as those Wilson coefficient combinations correspond to the lowest signal yields. %
The tighter 68\%~CL contour at $\cgghh\approx 0$ in Figure~\ref{sfg:gghh2d} can also be attributed to this excess.
As the signal yield is minimal around $\chhh = 3$, for higher \chhh the $-2\ln\Lambda$ value will decrease in the \lephad~SLT high-\mhh category, while it increases in the \hadhad high-\mhh category that does not feature an excess.
Hence, the slope of the combined $-2\ln\Lambda$ function is small in that region of the parameter space, leading to the observed behaviour.
The same effect can also be seen in Figure~\ref{sfg:klambdaObsExp}.

\FloatBarrier

\section{Conclusion}

\label{sec:concl}
An updated search for non-resonant Higgs boson pair production in the \bbtt final state has been performed using the full Run 2 ATLAS dataset, corresponding to 140~fb$^{-1}$ of proton--proton collisions at a centre-of-mass energy of $13$~TeV.
The results supersede and expand on those of a previous search based on the same dataset described in Ref.~\cite{HDBS-2018-40}.
Compared with the previous publication, the event classification has been reoptimised to enhance the sensitivity to \kl and to the VBF production mode.
Improved multivariate classifiers are used to build the final discriminants, increasing the sensitivity to SM \HH production and to anomalous values of the coupling modifiers \kl and \kvv.
The analysis sensitivity is improved by 11\% to 19\%, depending on the parameter of interest.
Results interpreted in terms of ggF and VBF production modes have been added, compared with the results of Ref.~\cite{HDBS-2018-40}.
The statistical procedure for the interpretation of the observed yields in terms of the signal coupling modifiers has been updated from scans of cross-section limits to profile likelihood ratio scans.

No evidence of \HH signal is found. An observed 95\% CL upper limit of 5.9 is set on the \HH production signal strength $\mu_{\HH}$, to be compared with an expected limit of 3.3 in the background-only ($\mu_{HH}=0$) hypothesis.
The observed limit on $\mu_{HH}$ is looser than the expected one as a result of a mild excess in the \lephad SLT SR, in the high-\mhh category.
The corresponding observed (expected) 95\% confidence intervals for the self-coupling modifier \kl and the quartic coupling modifier \kvv are $-3.1 < \kl < 9.0$ ($-2.5 < \kl < 9.3$) and $-0.5 < \kvv < 2.7$ ($-0.2 < \kvv < 2.4$), respectively.
Based on these improvements, 95\%~CL upper limits are set on the \HH production cross-section in the seven shape benchmark scenarios proposed in Ref.~\cite{alasfar:2023}.
Moreover, one-dimensional constraints are set on the Wilson coefficients \cH and \cHbox in the SMEFT framework as well as on \cgghh and \ctthh in the HEFT framework.


%
\section*{Acknowledgements}
%

%
%

%
%

We thank CERN for the very successful operation of the LHC and its injectors, as well as the support staff at
CERN and at our institutions worldwide without whom ATLAS could not be operated efficiently.

The crucial computing support from all WLCG partners is acknowledged gratefully, in particular from CERN, the ATLAS Tier-1 facilities at TRIUMF/SFU (Canada), NDGF (Denmark, Norway, Sweden), CC-IN2P3 (France), KIT/GridKA (Germany), INFN-CNAF (Italy), NL-T1 (Netherlands), PIC (Spain), RAL (UK) and BNL (USA), the Tier-2 facilities worldwide and large non-WLCG resource providers. Major contributors of computing resources are listed in Ref.~\cite{ATL-SOFT-PUB-2023-001}.

We gratefully acknowledge the support of ANPCyT, Argentina; YerPhI, Armenia; ARC, Australia; BMWFW and FWF, Austria; ANAS, Azerbaijan; CNPq and FAPESP, Brazil; NSERC, NRC and CFI, Canada; CERN; ANID, Chile; CAS, MOST and NSFC, China; Minciencias, Colombia; MEYS CR, Czech Republic; DNRF and DNSRC, Denmark; IN2P3-CNRS and CEA-DRF/IRFU, France; SRNSFG, Georgia; BMBF, HGF and MPG, Germany; GSRI, Greece; RGC and Hong Kong SAR, China; ISF and Benoziyo Center, Israel; INFN, Italy; MEXT and JSPS, Japan; CNRST, Morocco; NWO, Netherlands; RCN, Norway; MEiN, Poland; FCT, Portugal; MNE/IFA, Romania; MESTD, Serbia; MSSR, Slovakia; ARIS and MVZI, Slovenia; DSI/NRF, South Africa; MICINN, Spain; SRC and Wallenberg Foundation, Sweden; SERI, SNSF and Cantons of Bern and Geneva, Switzerland; NSTC, Taipei; TENMAK, T\"urkiye; STFC/UKRI, United Kingdom; DOE and NSF, United States of America.

Individual groups and members have received support from BCKDF, CANARIE, CRC and DRAC, Canada; CERN-CZ, PRIMUS 21/SCI/017 and UNCE SCI/013, Czech Republic; COST, ERC, ERDF, Horizon 2020, ICSC-NextGenerationEU and Marie Sk{\l}odowska-Curie Actions, European Union; Investissements d'Avenir Labex, Investissements d'Avenir Idex and ANR, France; DFG and AvH Foundation, Germany; Herakleitos, Thales and Aristeia programmes co-financed by EU-ESF and the Greek NSRF, Greece; BSF-NSF and MINERVA, Israel; Norwegian Financial Mechanism 2014-2021, Norway; NCN and NAWA, Poland; La Caixa Banking Foundation, CERCA Programme Generalitat de Catalunya and PROMETEO and GenT Programmes Generalitat Valenciana, Spain; G\"{o}ran Gustafssons Stiftelse, Sweden; The Royal Society and Leverhulme Trust, United Kingdom.

In addition, individual members wish to acknowledge support from CERN: European Organization for Nuclear Research (CERN PJAS); Chile: Agencia Nacional de Investigaci\'on y Desarrollo (FONDECYT 1190886, FONDECYT 1210400, FONDECYT 1230812, FONDECYT 1230987); China: National Natural Science Foundation of China (NSFC - 12175119, NSFC 12275265, NSFC-12075060); Czech Republic: PRIMUS Research Programme (PRIMUS/21/SCI/017); EU: H2020 European Research Council (ERC - 101002463); European Union: European Research Council (ERC - 948254, ERC 101089007), Horizon 2020 Framework Programme (MUCCA - CHIST-ERA-19-XAI-00), European Union, Future Artificial Intelligence Research (FAIR-NextGenerationEU PE00000013), Italian Center for High Performance Computing, Big Data and Quantum Computing (ICSC, NextGenerationEU); France: Agence Nationale de la Recherche (ANR-20-CE31-0013, ANR-21-CE31-0013, ANR-21-CE31-0022, ANR-22-EDIR-0002), Investissements d'Avenir Labex (ANR-11-LABX-0012); Germany: Baden-Württemberg Stiftung (BW Stiftung-Postdoc Eliteprogramme), Deutsche Forschungsgemeinschaft (DFG - 469666862, DFG - CR 312/5-2); Italy: Istituto Nazionale di Fisica Nucleare (ICSC, NextGenerationEU), Ministero dell'Università e della Ricerca (PRIN - 20223N7F8K - PNRR M4.C2.1.1); Japan: Japan Society for the Promotion of Science (JSPS KAKENHI JP21H05085, JSPS KAKENHI JP22H01227, JSPS KAKENHI JP22H04944, JSPS KAKENHI JP22KK0227); Netherlands: Netherlands Organisation for Scientific Research (NWO Veni 2020 - VI.Veni.202.179); Norway: Research Council of Norway (RCN-314472); Poland: Polish National Agency for Academic Exchange (PPN/PPO/2020/1/00002/U/00001), Polish National Science Centre (NCN 2021/42/E/ST2/00350, NCN OPUS nr 2022/47/B/ST2/03059, NCN UMO-2019/34/E/ST2/00393, UMO-2020/37/B/ST2/01043, UMO-2021/40/C/ST2/00187, UMO-2022/47/O/ST2/00148); Slovenia: Slovenian Research Agency (ARIS grant J1-3010); Spain: BBVA Foundation (LEO22-1-603), Generalitat Valenciana (Artemisa, FEDER, IDIFEDER/2018/048), La Caixa Banking Foundation (LCF/BQ/PI20/11760025), Ministry of Science and Innovation (MCIN \& NextGenEU PCI2022-135018-2, MICIN \& FEDER PID2021-125273NB, RYC2019-028510-I, RYC2020-030254-I, RYC2021-031273-I, RYC2022-038164-I), PROMETEO and GenT Programmes Generalitat Valenciana (CIDEGENT/2019/023, CIDEGENT/2019/027); Sweden: Swedish Research Council (VR 2018-00482, VR 2022-03845, VR 2022-04683, VR grant 2021-03651), Knut and Alice Wallenberg Foundation (KAW 2017.0100, KAW 2018.0157, KAW 2018.0458, KAW 2019.0447, KAW 2022.0358); Switzerland: Swiss National Science Foundation (SNSF - PCEFP2\_194658); United Kingdom: Leverhulme Trust (Leverhulme Trust RPG-2020-004); United States of America: U.S. Department of Energy (ECA DE-AC02-76SF00515), Neubauer Family Foundation.

%
%


%
%
%
%
%
%
%

%
\clearpage
\printbibliography
\clearpage
 
\begin{flushleft}
\hypersetup{urlcolor=black}
{\Large The ATLAS Collaboration}

\bigskip

\AtlasOrcid[0000-0002-6665-4934]{G.~Aad}$^\textrm{\scriptsize 103}$,
\AtlasOrcid[0000-0001-7616-1554]{E.~Aakvaag}$^\textrm{\scriptsize 16}$,
\AtlasOrcid[0000-0002-5888-2734]{B.~Abbott}$^\textrm{\scriptsize 121}$,
\AtlasOrcid[0000-0002-1002-1652]{K.~Abeling}$^\textrm{\scriptsize 55}$,
\AtlasOrcid[0000-0001-5763-2760]{N.J.~Abicht}$^\textrm{\scriptsize 49}$,
\AtlasOrcid[0000-0002-8496-9294]{S.H.~Abidi}$^\textrm{\scriptsize 29}$,
\AtlasOrcid[0009-0003-6578-220X]{M.~Aboelela}$^\textrm{\scriptsize 44}$,
\AtlasOrcid[0000-0002-9987-2292]{A.~Aboulhorma}$^\textrm{\scriptsize 35e}$,
\AtlasOrcid[0000-0001-5329-6640]{H.~Abramowicz}$^\textrm{\scriptsize 152}$,
\AtlasOrcid[0000-0002-1599-2896]{H.~Abreu}$^\textrm{\scriptsize 151}$
% ...
\bigskip
\\

% ...
$^{16}$Department for Physics and Technology, University of Bergen, Bergen; Norway.\\
% ...
$^{29}$Physics Department, Brookhaven National Laboratory, Upton NY; United States of America.\\
% ...
$^{35}$$^{(a)}$Facult\'e des Sciences Ain Chock, Universit\'e Hassan II de Casablanca;$^{(b)}$Facult\'{e} des Sciences, Universit\'{e} Ibn-Tofail, K\'{e}nitra;$^{(c)}$Facult\'e des Sciences Semlalia, Universit\'e Cadi Ayyad, LPHEA-Marrakech;$^{(d)}$LPMR, Facult\'e des Sciences, Universit\'e Mohamed Premier, Oujda;$^{(e)}$Facult\'e des sciences, Universit\'e Mohammed V, Rabat;$^{(f)}$Institute of Applied Physics, Mohammed VI Polytechnic University, Ben Guerir; Morocco.\\
% ...
$^{44}$Physics Department, Southern Methodist University, Dallas TX; United States of America.\\
% ...
$^{49}$Fakult\"{a}t Physik , Technische Universit{\"a}t Dortmund, Dortmund; Germany.\\
% ...
$^{55}$II. Physikalisches Institut, Georg-August-Universit\"{a}t G\"ottingen, G\"ottingen; Germany.\\
% ...
$^{103}$CPPM, Aix-Marseille Universit\'e, CNRS/IN2P3, Marseille; France.\\
% ...
$^{121}$Homer L. Dodge Department of Physics and Astronomy, University of Oklahoma, Norman OK; United States of America.\\
% ...
$^{151}$Department of Physics, Technion, Israel Institute of Technology, Haifa; Israel.\\
$^{152}$Raymond and Beverly Sackler School of Physics and Astronomy, Tel Aviv University, Tel Aviv; Israel.\\
% ...
$^{a}$ Also Affiliated with an institute covered by a cooperation agreement with CERN.\\
$^{b}$ Also at An-Najah National University, Nablus; Palestine.\\
$^{c}$ Also at Borough of Manhattan Community College, City University of New York, New York NY; United States of America.\\
$^{d}$ Also at Center for High Energy Physics, Peking University; China.\\
$^{e}$ Also at Center for Interdisciplinary Research and Innovation (CIRI-AUTH), Thessaloniki; Greece.\\
$^{f}$ Also at Centro Studi e Ricerche Enrico Fermi; Italy.\\
% ...

\end{flushleft}


%

%
%

%

%
%
%
%

\end{document}